\documentclass{dune}
\usepackage{comment}
\pdfoutput=1

\input{common/preamble}

\begin{document}

\nolinenumbers

\pagestyle{titlepage}


\pagestyle{titlepage}

\date{}

\title{\scshape\Large Snowmass Neutrino Frontier: \\
NF02 Topical Group Report\\
\normalsize Understanding Experimental Neutrino Anomalies
\vskip -10pt
\snowmasstitle
}


\renewcommand\Authfont{\scshape\small}
\renewcommand\Affilfont{\itshape\footnotesize}

\author[1]{G.~Karagiorgi}
\author[2]{B.~R.~Littlejohn}
\author[3]{P.~Machado}
\author[4]{A.~Sousa}
\author[ ]{\\ on behalf of the NF02 Topical Group Community$^{\thanks{This report is based on the NF02-contributed \emph{White Paper on Light Sterile Neutrino Searches and Related Phenomenology}~\cite{Acero:2022wqg}.}}$ }

\vspace{-0.5cm}
\affil[1]{Columbia University, New York, NY 10027, USA}
\affil[2]{Illinois Institute of Technology, Chicago, IL 60616, USA}
\affil[3]{Fermi National Accelerator Laboratory, Batavia, IL, USA}
\affil[4]{Department of Physics, University of Cincinnati, Cincinnati, OH 45221, USA}

\maketitle

\renewcommand{\familydefault}{\sfdefault}
\renewcommand{\thepage}{\roman{page}}
\setcounter{page}{0}

\pagestyle{plain} 
\clearpage
\textsf{\tableofcontents}





\renewcommand{\thepage}{\arabic{page}}
\setcounter{page}{1}

\pagestyle{fancy}

\fancyhead{}
\fancyhead[RO]{\textsf{\footnotesize \thepage}}
\fancyhead[LO]{\textsf{\footnotesize \nouppercase{\leftmark}}}

\fancyfoot{}
\fancyfoot[RO]{\textsf{\footnotesize Snowmass 2021}}
\fancyfoot[LO]{\textsf{\footnotesize NF02 Topical Group Report}}
\fancypagestyle{plain}{}

\renewcommand{\headrule}{\vspace{-4mm}\color[gray]{0.5}{\rule{\headwidth}{0.5pt}}}



\clearpage

\section*{Executive Summary}
\label{sec:summary}
\markboth{Executive Summary}{}

The past three decades of experimental neutrino measurements have accumulated observations of anomalous short-baseline flavor transformation from four different sectors of varied neutrino source (proton accelerators, reactors, and intense radioactive sources) and energy (from MeV to GeV scales).  Some of these observations have grown in significance over time to nearly $5\sigma$, and all share the commonality of having been observed at relatively short neutrino propagation distances, before any three-neutrino oscillation effects are expected to become measurable. They thus serve as an intensifying experimental impetus for pursuing beyond Standard Model (BSM) physics accessible through the neutrino sector, such as new neutrino mass states and hidden-sector couplings.  With the generation mechanism of neutrino mass currently unknown, these BSM searches are also theoretically well-motivated. 

Over the past ten years, the neutrino community has successfully implemented the primary recommendations of the previous P5 Report by developing and carrying out a diverse program of small-scale experiments aimed at directly addressing these short-baseline anomalies and at probing the leading (at the time) theoretical explanation for their existence: oscillations due to a single eV-scale, mostly-sterile neutrino state, or a ``3+1'' scenario.  
As part of that effort, recent reactor experiments have ruled out a large part of favored 3+1 oscillation parameter space, and uncovered evidence favoring an alternate conventional explanation for the Reactor Anomaly, namely limitations in our understanding of reactor fluxes. On the other hand, a direct test of the Gallium Anomaly was completed by the BEST experiment, providing a striking modern confirmation of this anomaly.  MiniBooNE Anomaly investigations have been augmented by further MiniBooNE data-taking and by MicroBooNE, revealing a complex picture seemingly at odds with a simple flavor transformation scenario as well as other popular conventional explanations.  
Long-anticipated new facilities with game-changing sensitivity to the LSND and MiniBooNE Anomalies---the Fermilab-based Short-Baseline Neutrino (SBN) Program and the JPARC-based JSNS$^2$---were initiated and are now online, promising powerful tests of leading interpretations of the anomalies.  
The community's accomplishments have also extended beyond the scope of the P5 recommendations by confronting the 3+1 sterile oscillation explanation with complementary datasets from MINOS/MINOS+, IceCube, T2K, NOvA, SuperK, KATRIN, and others, and by developing a rich array of additional new physics models---from exotic flavor transformation to hidden sector couplings---that represent viable interpretations of the anomalies.  

While not all anomalous observations have been resolved, 
the improvement in knowledge over the past decade has been significant, and compels deeper probes that are possible with current and upcoming experimental capabilities and facilities.  
Whether the anomalies are caused by a unified but complex regime of new physics, or by individual systematic or BSM effects specific to subsets of neutrino source types, the adoption of a multi-probe test strategy is essential to their convincing resolution.  
Thanks to joint efforts by the theoretical and experimental communities, the path forward to resolution of the anomalies is clearer, as are the possible range of exciting new physics scenarios that could potentially be responsible for them.  

A key ingredient to unearthing the origins of the short-baseline neutrino anomalies during the next decade is continued support for and full exploitation of short-baseline experiment investments made over the past decade.  
To that end, measurements by operating or imminent experiments will provide direct experimental tests of most of the anomalies and new information during the first half of the coming P5 period, which will enable tests of many popular interpretations with high sensitivity.  
This program represents a large international commitment with several involved experimental efforts sited at US National Laboratories.  
The Fermilab SBN Program can test a vast array of MiniBooNE Anomaly interpretations, ranging from conventional origins, to flavor transformation, to new particle production in the beam or in neutrino scattering.  
JSNS$^2$ at JPARC 
can test much of the suggested sterile oscillation parameter space from the LSND Anomaly, as well as many lepton flavor violation interpretations.  
Finally, existing compact reactor core facilities and scintillator detector technologies are well-positioned to quickly and inexpensively close out remaining regions of sterile oscillation space while more precisely measuring reactor fluxes at varied core types.  

Sustained community effort over the past decade has been expended to develop new experimental concepts including cutting-edge sources and/or techniques for probing the primary BSM-related anomaly origin categories---flavor transformation phenomena and new particle production---with significantly improved sensitivity over the aforementioned experimental efforts.  
In some cases, timely installation and completion of new experimental concepts can provide a new wave of valuable short-baseline information well before the close of the coming P5 period.  
Other concepts relying on facilities or detectors that can be available in the coming P5 period can enable this new wave to be sustained well into the following decade.
As an example of these new concepts, precise new BSM flavor transformation measurements can be made in the electron disappearance channel using precision measurements of beta decay and electron capture, novel isotope decay-at-rest sources, or large-volume short-baseline reactor or radioactive source experiments; in the muon disappearance channel using atmospheric neutrinos or neutrinos from decays of accelerator-produced kaons; and in a well-characterized mixture of \numu$\rightarrow$\nue appearance and/or \nue/\numu/all-active-flavor disappearance effects using future long-baseline oscillation facilities or coherent elastic neutrino-nucleus scattering-based measurements of decay-at-rest neutrinos.  
In addition, a vast array of potential dark sector couplings can be probed with high-statistics datasets from intense decay-in-flight, decay-at-rest, and collider beam dump neutrino sources or with atmospheric neutrino fluxes.  
Further, the community must also investigate ways of improving the understanding of the Gallium Anomaly, in conjunction with the potential interpretation of the other anomalies, and/or independently.
Finally, the continued investigations of complementary and/or indirect probes offered through currently-operating and future long-baseline accelerator measurements, atmospheric and solar neutrino constraints, precision measurements of weak nuclear decay products, neutrinoless double beta decay measurements, and constraints from cosmology will provide additional and important context for the ultimate resolution of the anomalies.  

Full exploitation of the above investments provides valuable new information and plausible paths to convergence for interpretation of the short-baseline anomalies, or potentially to exciting new physics discoveries. 
It is likely that these advancements will cause substantial shifts in the short-baseline experimental and theoretical landscape, potentially as early as by the middle of the upcoming P5 period, with subsequent priorities depending on what is learned about the various hypothesized anomaly origins.  
To prepare for this likelihood, the community should work presently to explore, develop, and support diverse opportunities for new dedicated measurements at existing or future facilities, and it should also be prepared to re-evaluate priorities given what will be learned.  
Beyond the community, it is also essential that available supporting resources are structured to enable and facilitate adaptation to developments during the next P5 period, and to ensure that timely experimental groundwork is laid for following generations of precision short-baseline measurements.



\section{Experimental Anomalies: Introduction and Motivation}
\label{sec:expt_landscape}

The Nobel prize-winning discovery of neutrino oscillation~\cite{Fukuda:1998mi,SNO:2001kpb,SNO:2002tuh} has led to a picture of three neutrino masses and flavor states that is now established as a minimal extension to the Standard Model (SM), and which is only empirically motivated. 
Neutrino mass generation is qualitatively different from that of any other fermions in the SM:  
in general, its mechanism would require the addition of particle content to the SM that has never been observed.  
The lack of experimental indication of the scale of this new physics makes the neutrino sector a promising portal to new physics.  
In particular, the existence of new states or gauge interactions associated with neutrinos could affect neutrino experiments in a variety of ways, such as effects on oscillation phenomenology or new particles produced in neutrino beams or in neutrino detectors.

Interest in this direction has been fanned by a series of anomalous experimental measurements, especially since the mid-1990's, which suggested the existence of new, light neutrino states. 
At the LSND experiment~\cite{Aguilar:2001ty}, an unexpected 3.8$\sigma$ excess of electron antineutrino events was observed from a meson decay-at-rest neutrino source, while at the MiniBooNE experiment~\cite{AguilarArevalo:2008yp}, a 4.8$\sigma$ excess of electron-anti/neutrino-like events was observed in a muon-anti/neutrino-pure beam from meson decay-in-flight.  
Additionally, many reactor neutrino experiments performed over many decades and at different baselines below 100~m observed a $\mathcal{O}$(5\%) deficit of electron antineutrino interactions with respect to theoretical predictions~\cite{Mention:2011rk}, while radiochemical experiments have observed a consistent $\mathcal{O}$(20\%) deficit in the electron neutrino rate expected from nearby intense electron capture radioactive sources~\cite{Hampel:1997fc, SAGE:1998fvr, Gavrin:2010qj}, with the most recent verification of this deficit provided by the BEST experiment, at $>4\sigma$ \cite{Barinov:2021asz}. This series of indications of neutrino phenomena deviating from the three-neutrino paradigm have commonalities of being observed primarily in electron (anti)neutrino observations, from either electron (anti)neutrino or muon (anti)neutrino sources, with either Cherenkov or scintillator detectors, and at relatively ``short baselines'' from these neutrino sources\footnote{Note that the term ``short baseline'' refers to a ratio of neutrino propagation distance relative to neutrino energy of $\sim1$~km/GeV or $\sim1$~m/MeV, corresponding to a neutrino oscillation frequency $\Delta$m$^2$ of $\sim1$~eV$^2$.}. 

These ``short-baseline experimental anomalies'', and the extensive and dedicated scientific program that has been launched over the past two decades to address them, is the focus of this Neutrino Frontier Topical Group, NF02. 
While a viable, experimentally-verified theoretical description of short-baseline experimental anomalies has not yet been identified, impending and future experiments promise to deliver new and invaluable information that can resolve the underlying origins of the anomalies. 
The anomalous experimental signatures, their potential interpretation(s), and past and future experimental efforts to address them, are described in detail in the Snowmass NF02-contributed White Paper~\cite{Acero:2022wqg}.  
The \textbf{ key questions } the NF02 community aims to address are the following:

\begin{itemize}

 \item \textit{What is the underlying physics behind the LSND, MiniBooNE, Reactor, and Gallium Anomalies?}

 \item \textit{If new physics is at play, what are the broader impacts within neutrino physics (long-baseline measurements, neutrinoless double-beta decay (0$\nu\beta\beta$) searches, neutrino mass measurements, neutrino interaction measurements) and beyond (cosmology, astrophysics, collider physics)?}

 \item \textit{If the resolution of the short-baseline anomalies lies in our inadequate understanding of standard particle processes, what is the impact of those processes, once understood, on future SM and BSM physics searches in neutrino experiments?}

\end{itemize}


\newpage
\section{Recent Progress: New Results, New Interpretations}
\label{sec:th_landscape}

Over the past ten years, the neutrino community has successfully implemented the primary recommendations of the previous P5 Report by developing and carrying out a diverse program of small-scale experiments aimed at directly addressing these short-baseline anomalies and at probing the leading theoretical explanation for their existence: light sterile neutrino oscillations.  
The community’s accomplishments have also extended beyond the scope of the P5 recommendations by confronting the sterile neutrino explanation with existing complementary datasets, and by developing a rich array of new physics models that represent viable interpretations of the anomalies.  
This Section will briefly summarize recent achievements, which are more thoroughly described in the NF02-wide White Paper~\cite{Acero:2022wqg}.

\paragraph{Recent Searches for eV-scale Sterile Neutrino Oscillations}\mbox{}\\
One of the most widely examined theoretical frameworks considered for the interpretation of these anomalies is the existence of oscillations mediated by one or more additional light ($\sim1$~eV) sterile neutrinos, referred to as a ``3+N'' oscillation model~\cite{Abazajian:2012ys}.  
This hypothesis is well-motivated by, for example, the anomalies found in LSND and MiniBooNE.  These experiments share a similar ratio between baseline and energy, $L/E$, even though the baselines and energies differ by an order of magnitude each; both observe $\nu_e$ or $\bar{\nu}_e$ excesses with event rates that are about 0.3\% of the measured muon neutrino event rates.  
By three-neutrino unitarity constraints~\cite{Parke:2015goa}, existing neutrino data can only allow for a small admixture of active neutrinos in the eV-scale mass eigenstate(s), which leads to possible short-baseline neutrino oscillations with appearance oscillation probabilities of order 1\% or less, and disappearance oscillation probabilities of up to a few tens of percent. Moreover, appearance and disappearance probabilities are correlated: a higher $\nu_\mu\rightarrow\nu_e$ appearance probability amplitude implies a higher product of $\nu_e$ and $\nu_\mu$ disappearance amplitudes.  

In the past ten years, a series of targeted short-baseline experiments has generated valuable new data regarding the existence of an additional eV-scale sterile neutrino state.  
In the electron-flavor disappearance channel, reactor-based experiments, including DANSS~\cite{DANSS:2018fnn}, NEOS~\cite{NEOS:2016wee},  PROSPECT~\cite{PROSPECT:2018dtt}, and STEREO~\cite{STEREO:2019ztb}, have observed no strong evidence for oscillations, limiting sin$^2 2\theta_{14}$ below 1-10\% amplitude over a wide mass splitting range from roughly 0.1 to 5~eV$^2$.  
More recently, the MicroBooNE experiment reported its first MiniBooNE-relevant beam $\nu_e$ measurement results, showing no strong initial indications of anomalous behavior~\cite{MicroBooNE:2021rmx,MicroBooNE:2021jwr,MicroBooNE:2021sne,MicroBooNE:2021nxr} and providing first tests of some LSND and MiniBooNE suggested $\nu_e$ appearance phase space regions~\cite{Denton:2021czb, Arguelles:2021meu, ub_osc_dl,ub_osc_wirecell}.  
In contrast, the BEST experiment reproduced an anomalous deficit in radioactive source $\nu_e$ interaction rates matching that observed by the previous SAGE and GALLEX experiments~\cite{Hampel:1997fc, SAGE:1998fvr, Gavrin:2010qj}.  
Beyond targeted short-baseline efforts, the IceCube\footnote{While IceCube provides an exclusion region at 99\%~CL, it observes a closed 90\%~CL contour consistent with oscillations~\cite{IceCube:2020tka}.}~\cite{IceCube:2020phf,IceCube:2020tka,IceCube:2017ivd,Aartsen:2017mnf} and MINOS/MINOS+~\cite{MINOS:2017cae, MINOS:2016viw,MINOS:2011ysd} experiments placed new, tight limits on $\nu_{\mu}$ disappearance and $\theta_{24}$ in the eV-scale regime, the new KATRIN neutrino mass experiment constrained sin$^2 2\theta_{14}$ above 10~eV$^2$~\cite{KATRIN:2020dpx}, and sterile searches at Super Kamiokande, MINOS/MINOS+, IceCube, and NOvA yielded first limits on $\theta_{34}$~\cite{Super-Kamiokande:2014ndf,MINOS:2017cae,IceCube:2017ivd,NOvA:2017geg,NOvA:2021smv}.  
While the 3+$N$ sterile neutrino oscillation framework could simultaneously accommodate all short-baseline experimental anomalies on their own, it fails when facing the enhanced scrutiny of multiple other measurements, and much more spectacularly with today's accumulated higher-precision measurements.  
Specifically, the relatively large active-sterile mixing needed to interpret MiniBooNE and LSND data cannot be accommodated in the face of strong limits on $\nu_e$ and $\nu_{\mu}$ disappearance from reactor and accelerator/atmospheric experiments, respectively.  
Moreover, the BEST-affirmed Gallium Anomaly requires relatively large ($>$10\%) $\nu_e$ disappearance probabilities, which are disfavored \cite{Barinov:2021asz} by data from reactor and solar neutrino experiments, and is in strong tension with cosmological measurements of the sum of neutrino masses~\cite{Planck:2018vyg}.

Especially in the past five years, these contradictions spurred the community to look beyond 3+$N$ sterile neutrino models for explanations to all or part of the anomalies.  
This broader line of theory development can be categorized into three classes: flavor conversion models; dark sectors in scattering and in the beam; and conventional (but not well understood) SM origins.  

\paragraph{Flavor Conversion Model Development}\mbox{}\\
This class of theories includes models with one (3+1) or more (3+$N$) light sterile neutrinos, possibly with novel dynamics that lead to non-standard interactions affecting oscillation physics~\cite{Akhmedov:2011zza,Bramante:2011uu,Karagiorgi:2012kw,Asaadi:2017bhx,Smirnov:2021zgn,Alves:2022vgn} or that lead to decays of these light states~\cite{Palomares-Ruiz:2005zbh,Bai:2015ztj,Moss:2017pur,Moulai:2019gpi,deGouvea:2019qre,Dentler:2019dhz,Hostert:2020oui}; scenarios in which neutrinos propagate through extra dimensions, changing their oscillation pattern and/or their dispersion relation; violations of Lorentz symmetry; new physics modifying muon decays~\cite{Bergmann:1998ft,Babu:2016fdt,Jones:2019tow}; and others~\cite{Berryman:2018ogk}.
In general, the smoking gun signatures of such scenarios lie in the realm of neutrino flavor  phenomenology, many times presenting complex $L/E$ dependent effects, as well as nontrivial correlations among different oscillation channels.  
Complementary measurements across disparate oscillation channels, neutrino sources, neutrino energies, and experimental baselines is particularly important in distinguishing between the different flavor transition model candidates.

\paragraph{Dark Sector Scattering Scenarios}\mbox{}\\
These models are inspired by the theoretical observation that neutrinos are excellent candidates to be portals to dark sectors.  
In these scenarios, new particles are created either in the beam, by the decay of mesons, or by the neutrino interaction in the detector.  
The decays of the new particles lead to events that would be interpreted as $\nu_e$- or $\bar\nu_e$-like, and thus could explain the excess observed in LSND and/or MiniBooNE.  
Models that fit this category include transition magnetic moment scenarios~\cite{Gninenko:2009ks,Gninenko:2010pr,Gninenko:2012rw,Masip:2012ke,Radionov:2013mca,Magill:2018jla,Vergani:2021tgc,Alvarez-Ruso:2021dna}, dark neutrinos~\cite{Bertuzzo:2018itn,Bertuzzo:2018ftf,Ballett:2018ynz,Ballett:2019pyw,Datta:2020auq,Dutta:2020scq,Abdullahi:2020nyr,Abdallah:2020biq,Abdallah:2020vgg,Hammad:2021mpl}, dark scalars~\cite{Chang:2021myh}, dark matter scenarios~\cite{Dutta:2021cip}, and heavy neutral leptons~\cite{Fischer:2019fbw}.  

The experimental signatures of these models are rich and varied when compared to flavor conversion scenarios. 
For example, in transition magnetic moment scenarios, neutrinos up-scatter to heavy neutral fermions, which decay to a neutrino and a photon, leading to an overall signature with some hadronic activity and a final state photon.  
In contrast, dark neutrinos up-scatter to a new particle via a new gauge boson, followed by a decay to a neutrino and a potentially collimated $e^+e^-$ pair (either via two- or three-body decays).  
While in MiniBooNE all these models would lead to similar observables explaining its anomalous results, in other experiments, such as MicroBooNE and other liquid argon time projection chambers (LArTPCs), signatures differ in several respects: the presence or absence of hadronic activity, which affects vertex reconstruction; a possible gap between the vertex and the electromagnetic shower inconsistent with the photon conversion length in argon; different ionization yields in the beginning of the electromagnetic shower; or potentially a noticeable opening angle for the $e^+e^-$ pair.

\paragraph{``Conventional'' Anomaly Origins}\mbox{}\\ It is also possible that some or all of the anomalies have conventional rather than BSM origins.  
While in this case each experimental anomaly would likely have to be explained by a different, unrelated origin, this is a viable possibility given the complexities in modelling signatures from each relevant experiment type.  
Improper modelling of antineutrino production~\cite{Hayes:2013wra,bib:IAEA,Hayen:2019eop,fijalkowska2017impact,Guadilla:2019gws,Guadilla:2019zwz,IGISOL:2015ifm,rasco2016decays,Rice:2017kfj,Valencia:2016rlr,Hayes:2017res,Estienne:2019ujo} has been explored in detail and stands as a likely explanation for part, if not all, of the Reactor Anomaly. Uncertainties in neutrino-nucleus interaction cross sections~\cite{Bahcall:1997eg,Giunti:2010zu,Frekers:2011zz,Frekers:2013hsa,Frekers:2015wga,Honma:2009zz,Kostensalo:2019vmv} have been studied as a potential explanation for the Gallium Anomaly, though current understanding indicates improved cross section calculations might only reduce the significance of the anomaly by up to $1\sigma$~\cite{Berryman:2021yan}.
The MiniBooNE excess was measured amidst a substantial background contribution dominated by photon-producing neutrino-nucleus interactions~\cite{Hill:2009ek,Serot:2012rd,Wang:2013wva}, 
which have not been precisely measured and may carry substantial and poorly-defined theoretical uncertainties~\cite{Zhang:2012xn,Wang:2014nat,Giunti:2019sag}.  
Recent work has begun to probe these possibilities, with MicroBooNE experimentally exploring enhanced photon production via formation of $\Delta(1232)$ resonances~\cite{MicroBooNE:2021zai}, and ruling out the possibility of its radiative decay as a source of the MiniBooNE excess at $>95\%$~CL.

\paragraph{Summary of Recent Progress}\mbox{}\\
A summary of the anomaly interpretations detailed above is provided in Table~\ref{tab:sec-3:big_picture}, mapping specific models onto broader model classes and onto which anomalies they are meant to address.  
Table~\ref{fig:sec-3:prospects_big_picture} provides an orthogonal mapping between broad anomaly classes and experimental efforts capable of observing their relevant signatures.  
This latter Table includes entries for upcoming and proposed experiments alongside those of the recently-performed experiments described earlier.

\renewcommand{\arraystretch}{1}
\newcommand{\cmark}{\ding{51}}%
\newcommand{\xmark}{\ding{55}}%
\newcommand{\nocheck}{\textcolor{Red}{\xmark}}
\newcommand{\semicheck}{\textcolor{Orange}{\cmark}}
\newcommand{\fullcheck}{\textcolor{Green}{\cmark}}
\newcommand{\newhline}{\cline{2-8}}

\begin{table*}[!htp]
    \centering
    \begin{sideways}
\resizebox{20cm}{!}{%
    \begin{tabular}{|>{\centering\arraybackslash}p{3 cm}|>{\centering\arraybackslash}p{4 cm}|>{\centering\arraybackslash}p{3.2 cm}|>{\centering\arraybackslash}p{2 cm}|>{\centering\arraybackslash}p{1.9 cm}|>{\centering\arraybackslash}p{1.4 cm}|>{\centering\arraybackslash}p{1.3 cm}|>{\centering\arraybackslash}p{2.5 cm}|}
\hline\hline
    \multirow{2}{*}{Category} & \multirow{2}{*}{Model}& \multirow{2}{*}{Signature} & \multicolumn{4}{c|}{Anomalies} & \multirow{2}{*}{References} \\ \cline{4-7} & & & LSND & MiniBooNE & Reactor & Gallium & \\
\hline\hline
\multirow[c]{3}{*}[-3em]{ \parbox[t]{3 cm}{\vspace{-0.9cm}\centering \textbf{Flavor Conversion:} Transitions }}
        & 3+$N$ oscillations & oscillations & \fullcheck & \fullcheck & \fullcheck & \fullcheck & Reviews and global fits \cite{Dentler:2018sju,Diaz:2019fwt,Boser:2019rta,Dasgupta:2021ies}
        \\
        \newhline
        & 3+$N$ w/ invisible sterile decay & oscillations w/ $\nu_4$ invisible decay &  \fullcheck &  \fullcheck & \fullcheck & \fullcheck & \cite{Moss:2017pur,Moulai:2019gpi}
        \\
        \newhline
        & 3+$N$ w/ sterile decay & $\nu_4 \to \phi \nu_e$ &  \fullcheck &  \fullcheck & \semicheck & \semicheck & \cite{Palomares-Ruiz:2005zbh,Bai:2015ztj,deGouvea:2019qre,Dentler:2019dhz,Hostert:2020oui}
        \\
\hline\hline
\multirow[c]{2}{*}[-2em]{ \parbox[t]{3 cm}{\vspace{-0.5cm}\centering \textbf{Flavor Conversion:} Matter Effects }}
        & 3+$N$ w/ anomalous matter effects  & $\nu_\mu \to \nu_e$ via matter effects & \fullcheck  & \fullcheck & \nocheck & \nocheck & \cite{Akhmedov:2011zza,Bramante:2011uu,Karagiorgi:2012kw,Asaadi:2017bhx,Smirnov:2021zgn}
        \\ 
        \newhline
        & 3+$N$ w/ quasi-sterile neutrinos  & $\nu_\mu \to \nu_e$ w/ resonant $\nu_s$ matter effects & \fullcheck & \fullcheck & \semicheck & \semicheck & \cite{Alves:2022vgn}
        \\
\hline\hline
\multirow[c]{2}{*}[-2em]{ \parbox[t]{3 cm}{\vspace{-0.5cm}\centering \textbf{Flavor Conversion:} Flavor Violation  }} 
        & lepton-flavor-violating $\mu$ decays & $\mu^+\to e^+ \nu_{\alpha}\bar{\nu}_e$ & \fullcheck & \nocheck & \nocheck & \nocheck & \cite{Bergmann:1998ft,Babu:2016fdt,Jones:2019tow}
        \\ 
        \newhline
        & neutrino-flavor-changing bremsstrahlung & $\nu_\mu A \to e \phi A$ & \fullcheck & \fullcheck & \nocheck & \nocheck  & \cite{Berryman:2018ogk} 
        \\
\hline\hline
\multirow[c]{2}{*}[-2em]{ \parbox[t]{3 cm}{\vspace{-0.7cm}\centering \textbf{Dark Sector:} Decays in Flight }}
        & transition magnetic mom., heavy $\nu$ decay & $N\to \nu \gamma$ &  \nocheck & \fullcheck & \nocheck & \nocheck & \cite{Fischer:2019fbw}
        \\
        \newhline
        & dark sector heavy neutrino decay & $N\to \nu (X \to e^+e^-)$ or $N\to \nu (X\to \gamma \gamma)$ & \nocheck & \fullcheck & \nocheck & \nocheck & \cite{Chang:2021myh}
        \\
        \newhline
\hline\hline
\multirow[c]{2}{*}[-2em]{\parbox[t]{3 cm}{\vspace{-0.7cm}\centering \textbf{Dark Sector:} Neutrino Scattering }} 
    & neutrino-induced up-scattering & $\nu A \to N A$,  $N\to \nu e^+e^-$ or $N \to \nu \gamma \gamma$ & \semicheck & \fullcheck & \nocheck  & \nocheck & \cite{Bertuzzo:2018itn,Bertuzzo:2018ftf,Ballett:2018ynz,Ballett:2019pyw,Datta:2020auq,Dutta:2020scq,Abdullahi:2020nyr,Abdallah:2020biq,Abdallah:2020vgg,Hammad:2021mpl}
    \\
    \newhline
    & neutrino dipole up-scattering & $\nu A \to N A$,  $N\to \nu \gamma $ & \semicheck & \fullcheck & \nocheck & \nocheck  & \cite{Gninenko:2009ks,Gninenko:2010pr,Gninenko:2012rw,Masip:2012ke,Radionov:2013mca,Magill:2018jla,Vergani:2021tgc,Alvarez-Ruso:2021dna}
    \\
\hline\hline
\multirow[c]{2}{*}[-2em]{ \parbox[t]{3 cm}{\vspace{-0.9cm}\centering \textbf{Dark Sector:} Dark Matter Scattering }}
    & dark particle-induced up-scattering  & $\gamma$ or $e^+e^-$ & \nocheck & \fullcheck & \nocheck & \nocheck & \cite{Dutta:2021cip}
    \\
    \newhline
    & dark particle-induced inverse Primakoff & $\gamma$ & \fullcheck & \fullcheck & \nocheck & \nocheck & \cite{Dutta:2021cip}
    \\
\hline\hline
\end{tabular}
}
 \end{sideways}
    \caption{New physics explanations of the short-baseline anomalies categorized by their signature. Notation: \fullcheck -- the model can naturally explain the anomaly,  \semicheck -- the model can partially explain the anomaly,   \nocheck -- the model cannot explain the anomaly.  Table from Ref.~\cite{Acero:2022wqg}, with minor modifications.~\label{tab:sec-3:big_picture}}
\end{table*}



\renewcommand{\arraystretch}{1.2}
\newcommand{\fixedcol}{{\centering\arraybackslash}p{2.6 cm}}
\begin{table}[!htp]
\centering
\begin{sideways}
\resizebox{22cm}{!}{%
\begin{tabular}{
|>{\centering\arraybackslash}p{2.4 cm}
|>{\centering\arraybackslash}p{4.8 cm}
|>{\centering\arraybackslash}p{2.4 cm}
|>{\centering\arraybackslash}p{2.4 cm}
|>{\centering\arraybackslash}p{2.4 cm}
|>{\centering\arraybackslash}p{2.4 cm}
|>{\centering\arraybackslash}p{2.4 cm}|
            }
\hline \hline
& \textbf{Flavor Conversion: }
& \textbf{Flavor Conversion: }
& \textbf{Flavor Conversion: }
& \textbf{Dark Sector:}
& \textbf{Dark Sector:} 
& \textbf{Dark Sector:}
\\
Source 
& 3+$N$ Oscillations 
& Anomalous Matter Effects 
& Lepton Flavor Violation 
& Decays in Flight 
& Neutrino-induced Up-scattering 
& Dark-particle-induced Up-scattering 
\\  \hline \hline
Reactor  
        & DANSS Upgrade, JUNO-TAO, NEOS-II, Neutrino-4 Upgrade, PROSPECT-II & & & & & \\ 
\hline   
Radioactive Source
      & BEST-2, IsoDAR, THEIA, Jinping & & & & & \\  
\hline
Atmospheric
        & \multicolumn{2}{|>{\centering\arraybackslash}p{7.2 cm}|}{IceCube Upgrade, KM3NET, ORCA and ARCA, DUNE, Hyper-Kamiokande, THEIA} & & &  \multicolumn{2}{|>{\centering\arraybackslash}p{4.8 cm}|}{IceCube Upgrade, KM3NET, ORCA and ARCA, DUNE, Hyper-Kamiokande, THEIA}\\ 
\hline
Pion/Kaon Decay-At-Rest
        & JSNS$^2$, COHERENT, Coherent CAPTAIN-Mills, KPIPE & & JSNS$^2$, COHERENT, Coherent CAPTAIN-Mills, KPIPE, PIP2-BD & & & COHERENT, Coherent CAPTAIN-Mills, KPIPE, PIP2-BD, SBN-BD \\ 
\hline  
Beam Short Baseline
        & SBN & & & \multicolumn{3}{|>{\centering\arraybackslash}p{7.2 cm}|}{SBN, FASER$\nu$, SND@LHC, FLArE} \\  
\hline
Beam Long Baseline
        & \multicolumn{3}{|>{\centering\arraybackslash}p{9.6 cm}|}{DUNE, Hyper-Kamiokande, ESSnuSB} &  \multicolumn{3}{|>{\centering\arraybackslash}p{7.2 cm}|}{DUNE, Hyper-Kamiokande, ESSnuSB} \\  
\hline
Muon Decay-In-Flight 
      &  \multicolumn{3}{|>{\centering\arraybackslash}p{9.6 cm}|}{nuSTORM} & & nuSTORM &\\  
\hline
Beta Decay and Electron Capture  & KATRIN/TRISTAN, Project-8, HUNTER, BeEST, DUNE ($^{39}$Ar), PTOLEMY, $2\nu\beta\beta$ &  & & & & \\  
\hline\hline
\end{tabular}
}
\end{sideways}
\caption{Summary of current and future experimental prospects to probe new physics explanations of the anomalies. All experiments can constrain new physics models in direct and indirect ways; this table categorizes them according to their capability of providing a direct test of the respective model.  Table from Ref.~\cite{Acero:2022wqg}, with minor modifications.\label{fig:sec-3:prospects_big_picture}}
\footnotesize
\vspace{0.2in}
\end{table}


\section{The Path Towards Resolution of the Anomalies}
\label{sec:future}

While the underlying source of the anomalies remains unresolved, the improvement in knowledge over the past decade is significant, and compels deeper probes that are possible with current and upcoming experimental capabilities and facilities.  
Whether the anomalies are caused by a unified but complex regime of new physics, or by individual systematic or BSM effects specific to subsets of experiments, the adoption of a multi-probe test strategy is essential to their convincing resolution.  
Thanks to joint efforts by the theoretical and experimental communities, the path forward to resolution of the anomalies is clearer, as are the possible range of exciting new physics scenarios that could potentially be responsible for them.  

\subsection{Full Exploitation of Ongoing Experimental Programs}

A key ingredient to unearthing the origins of the short-baseline neutrino anomalies in the next decade is continued support for and full exploitation of short-baseline experiment investments made over the past decade.  
To that end, measurements by operating or imminent experiments at reactors, decay-at-rest, and decay-in-flight sources will provide direct experimental tests of most of the anomalies %
and new information during the first half of the coming P5 period which will 
enable new tests of many popular interpretations with high sensitivity.  
This program represents a large international commitment, with many of the involved experimental efforts sited at US National Laboratories.  
Full exploitation of these investments provides a promising path to convergence for interpretation of the LSND, MiniBooNE, and Reactor Anomalies.  

\begin{figure}[htbp!]
    \centering
    \includegraphics[width=0.52 \textwidth]{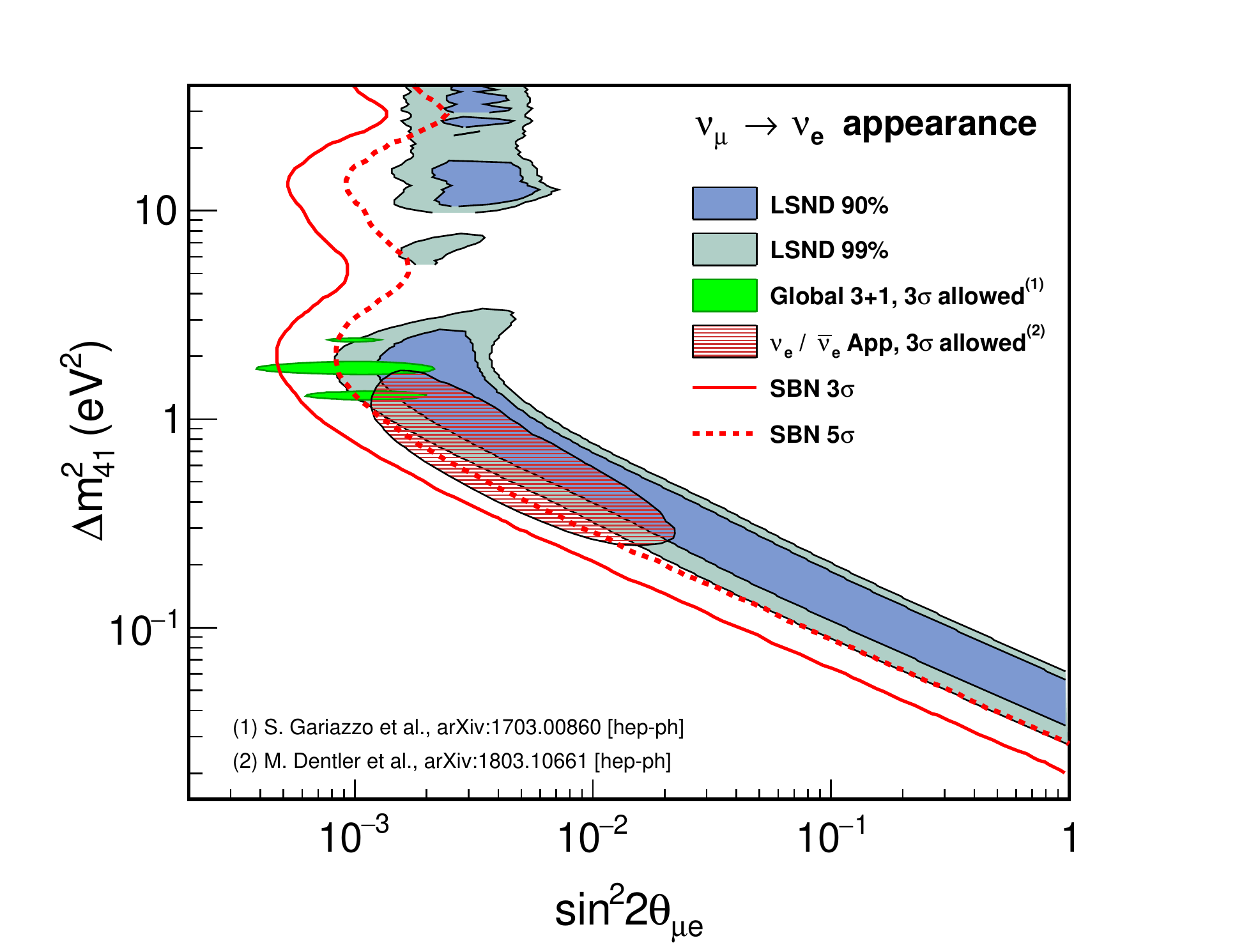}
    \includegraphics[width=0.46 \textwidth]{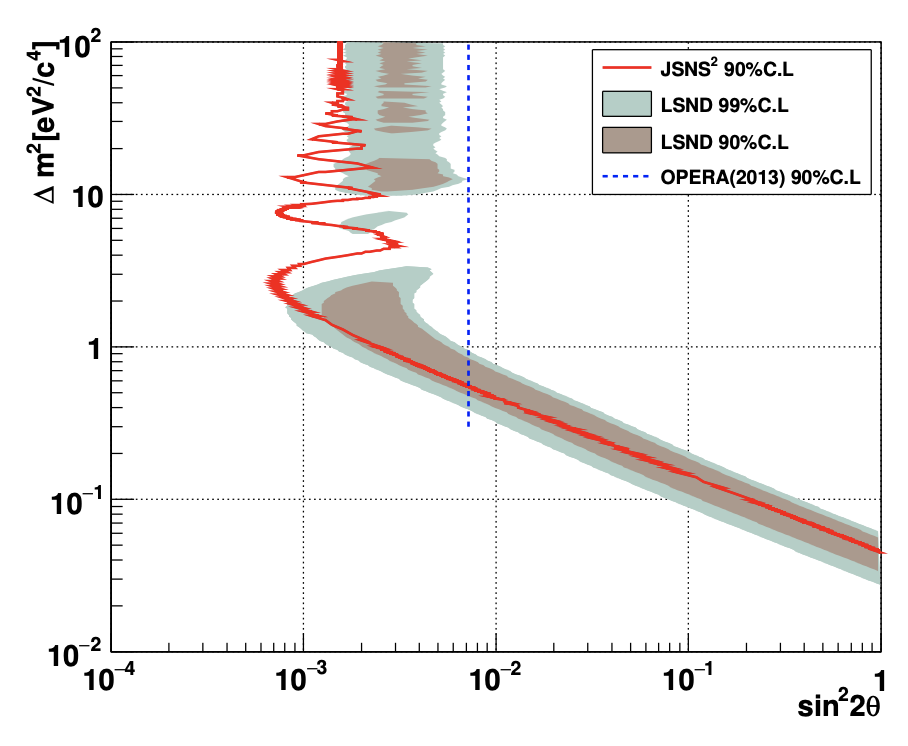}
    \caption{
      Expected light sterile neutrino oscillation sensitivities in the $\nu_\mu\to\nu_e$ appearance channel provided by the current DIF-based Fermilab SBN (left) and DAR-based JSNS$^2$ (right) experiments.  
      For SBN (JSNS$^2$) The $3\sigma$ and $5\sigma$ (90\% C.L.) sensitivities are given by the solid and dotted (solid) red curve.  LSND 90\% C.L. and 99\% C.L. allowed regions are also included~\cite{Aguilar:2001ty}.  For SBN, global $3\sigma$ $\nu_e$ appearance regions from Ref.~\cite{Dentler:2018sju} and global best-fit regions from Ref.~\cite{Gariazzo:2017fdh} are also shown.
    } 
    \label{FIG:SBN}
  \end{figure}

Meson decay-in-flight (DIF) beamlines offer direct access to flavor transformation, particle production, and conventional origin explanations of the MiniBooNE Anomaly.  
Building on the recent successes of the MicroBooNE LArTPC experiment~\cite{MicroBooNE:2016pwy}, the Fermilab SBN program \cite{MicroBooNE:2015bmn} will serve as the globe's pre-eminent dedicated short-baseline accelerator-based neutrino program for most of the next decade. 
The SBN program consists of three LArTPC detectors located along the BNB at Fermilab:  MicroBooNE, which completed operations in 2021; ICARUS, which began operations in the BNB in 2021; and the Short-Baseline Near Detector (SBND), which will begin operations in 2023. 
SBN will provide a multi-baseline search for anomalous flavor transformation in multiple exclusive and inclusive channels~\cite{Machado:2019oxb}---$\nu_\mu$ disappearance,  $\nu_\mu\rightarrow\nu_e$ appearance and/or $\nu_e$ disappearance, as well as neutral current $\nu_x\rightarrow\nu_s$ searches~\cite{Furmanski:2020smg}---that will enable decisive testing of the 3+$N$ sterile oscillation phase space as well as other broader classes of flavor transformation models~\cite{Cianci:2017okw}.  
The reach of SBN in probing LSND-related $\nu_\mu\rightarrow\nu_e$ phase space is demonstrated in Figure~\ref{FIG:SBN}. By its completion, SBN will probe the entirety of the relevant $\nu_\mu\rightarrow\nu_e$ space suggested by LSND and MiniBooNE Anomalies at high (3$\sigma$ or better) confidence level.
SBN's main physics goals also include detailed studies of neutrino–argon interactions at the GeV energy scale, including rare photon production processes that could serve as a more conventional origin of the MiniBooNE Anomaly.  
High statistics, coupled with the excellent LArTPC imaging and reconstruction capabilities of the SBN detectors, opens up invaluable opportunities for many of the dark sector searches arrayed in Tables~\ref{tab:sec-3:big_picture} and~\ref{fig:sec-3:prospects_big_picture} characterized by unique final-state signatures, such as $e^-, \, e^+e^-, \,  \gamma,$ $\gamma\gamma$, unusual hadronic content or topologies, and more.  

Meson and muon decay-at-rest (DAR) experiments offer a variety of unique attributes with respect to meson DIF experiments: in particular, they sample sources with extremely low $\bar{\nu}_e$ content and comparatively lower neutrino energies, and many use proton beamlines with high duty factors.  
The JSNS$^2$ experiment~\cite{JSNS2:2013jdh,Ajimura:2017fld,JSNS2:2021hyk}, currently the world's sole DAR-focused short-baseline effort aimed at directly measuring LSND's $\bar{\nu}_e$ excess, plans to use the same inverse beta decay (IBD) detection channel as LSND, but with an improved neutrino source and detector technology.  
Based at the 1~MW JPARC MLF proton beam and operating since 2021, JSNS$^2$ is taking data with a 17~t Gd-doped liquid scintillator detector placed 24~m from the beam target at a far off-axis angle.  
As shown in Figure~\ref{FIG:SBN}, JSNS$^2$ will be capable of testing the existence of an LSND-sized excess in their data, as well as LSND's suggested 3+1 oscillation phase-space, at high confidence level.  
With deployment of a functionally identical second detector at 48~m in the next few years, the proposed JSNS$^2$-II~\cite{Ajimura:2020qni} upgrade will substantially improve upon the effort's impending first results.  

\begin{figure}[ht]
    \centering
    \includegraphics[width=0.65 \textwidth,trim={0 0 0 4.0cm },clip]{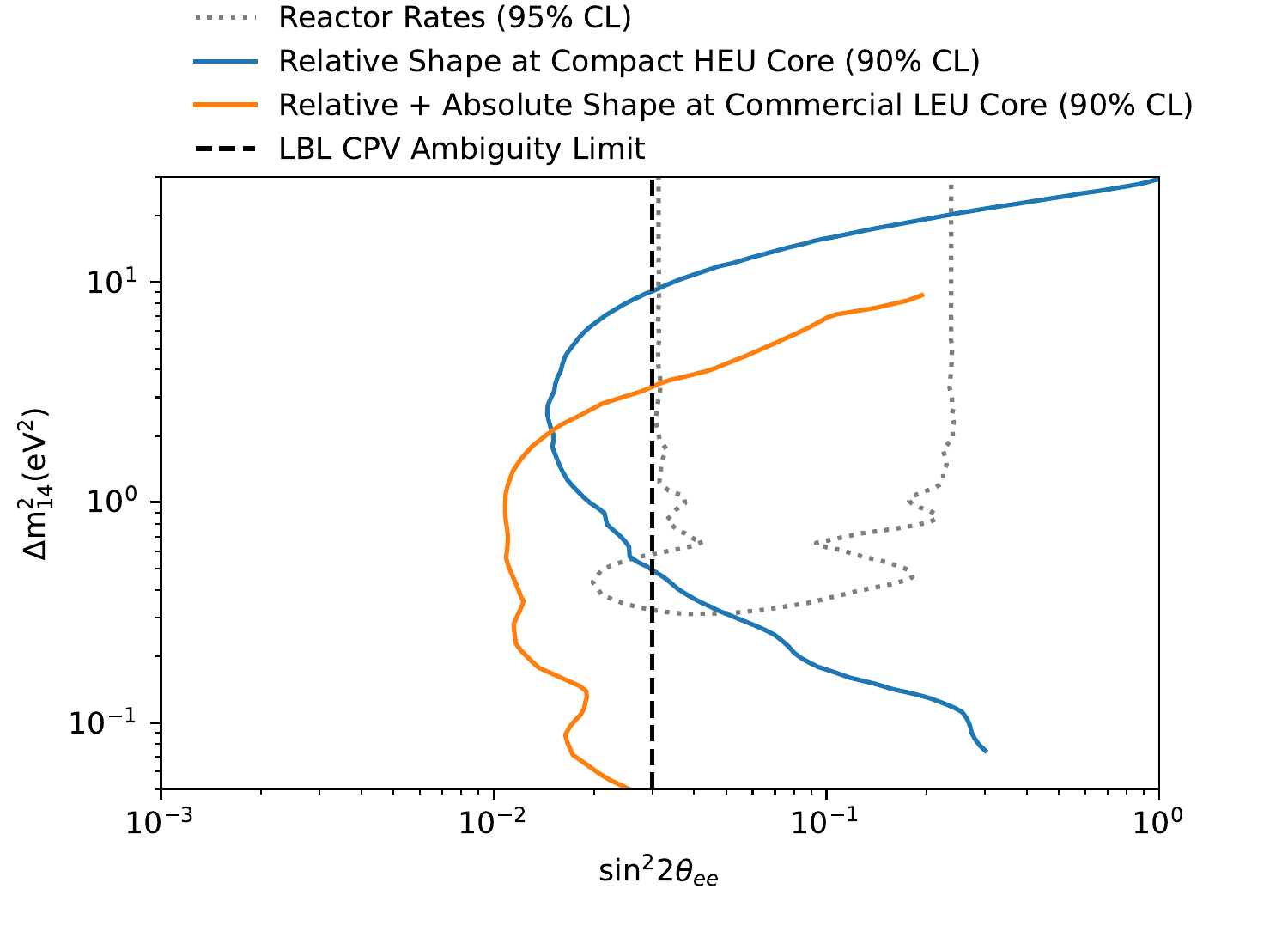}
    \caption{Example projected sensitivities~(90\% CL) to sterile neutrinos for future measurements performed at compact-core (blue, from Ref.~\cite{Andriamirado:2021qjc}) and larger commercial (orange, from Ref.~\cite{JUNO:2020ijm}) reactor cores.  Also shown is the region of phase space suggested by the Reactor Anomaly~\cite{Giunti:2021dkab}, and the active-sterile mixing sensitivity benchmark from Ref.~\cite{Gandhi:2015xza} required for clarity in interpreting long-baseline CP-violation results.  
}
    \label{fig:rx_future}
\end{figure}

Reactor experiments are unique among realized short-baseline efforts in their capability to cleanly probe sterile oscillation effects independently of most other phenomenological scenarios.  
This is due to the lower neutrino energies involved, which prohibits production and decay of heavier hidden-sector particles, their very short baselines, which minimize the impact of non-standard interactions during propagation, and their pure flavor content, which precludes ambiguities related to competing appearance and disappearance effects~\cite{Arguelles:2021meu}.  
While reactor experiments in the past decade have provided strong limits in oscillation phase space below a few~eV$^2$, limits above this mass splitting are substantially weaker; moreover, a claim of observation of a non-zero oscillation effect has also recently been made in this region by the Neutrino-4 experiment~\cite{Serebrov:2020kmd}, although with the caveats pointed out in Refs.~\cite{PROSPECT:2020raz,DanilovComment,Giunti:2021iti}.  
To address these issues at reactors, very short ($<$10~m) baseline measurements using compact cores and segmented detectors are required.  
Among this category, PROSPECT has set new oscillation limits at high $\Delta$m$^2$~\cite{PROSPECT:2018dtt,PROSPECT:2020sxr} and plans to use a second physics run to improve its statistical precision by more than an order of magnitude~\cite{Andriamirado:2021qjc}, while the CHANDLER collaboration has achieved on-surface detection of reactor \anue in a prototype detector demonstrating its novel plastic scintillator technology~\cite{Haghighat:2018mve}.  
An example 3+1 oscillation phase space coverage of such an experiment using reactor-model-independent analysis techniques is represented in Figure~\ref{fig:rx_future}, alongside an example from a future experiment at a larger commercial core.  
In addition to probing the remainder of low-$\Delta$m$^2$ 3+1 phase space suggested by the Reactor Anomaly, a near-term compact core experiment, combined with existing datasets, will generate few-percent-level active-sterile coupling sensitivity in the electron disappearance channel for all $\Delta$m$^2$ below roughly 10~eV$^2$, enabling high-confidence tests of oscillation claims from Neutrino-4 and greater clarity in interpretation of CP-violation results from DUNE~\cite{Gandhi:2015xza}.  
For more robust probes of absolute reactor neutrino fluxes and spectra, and thus better testing of conventional Reactor Anomaly origins, high-statistics datasets are needed from sources of widely varying fuel content, such as HEU reactors (the compact-core experiments described above) and full LEU reactor cycles (JUNO-TAO~\cite{JUNO:2020ijm},  NEOS-II~\cite{young_ju_ko_2020_3959599}, an upgraded DANSS experiment~\cite{svirida:2020danssup}, or a future PROSPECT-II LEU deployment~\cite{Andriamirado:2021qjc}).  
Quick timelines for achieving deployment ($<1$~year) and final results ($<$5 years) for these reactor efforts are possible due to the existence of vetted scintillator detector technologies and previously-qualified reactor sites (such as the ORNL-based HFIR reactor~\cite{PROSPECT:2015eri, PROSPECT:2020vcl}) from recently completed reactor experiment phases.

The combination of the results obtained by GALLEX~\cite{Hampel:1997fc}, SAGE~\cite{SAGE:1998fvr} and BEST~\cite{Barinov:2021asz} leads to a Gallium Anomaly with large significance, allowing the possibility that electron neutrinos may disappear through oscillations with the participation of a sterile neutrino at the $\mathcal{O}(\rm{eV})$ mass scale, that hidden sector processes are exerting their influence at the point of neutrino production or interaction, or that current nuclear modelling of neutrino-gallium interactions is inaccurate.  
While a convincing understanding of this anomaly still appears out of reach, there are few proposed experimental remedies capable of providing new direct insights on short timescales.  
Most prominently, the BEST Collaboration has proposed using their existing experimental configuration with a new $^{65}\rm{Zn}$ source~\cite{Gavrin:2018zmf,Gavrin:2019rtr}, which will result primarily in the detection of 1.35~MeV neutrinos, nearly twice as high in energy as its previous $^{51}\rm{Cr}$ source runs \cite{Bahcall:1997eg}; this primary difference offers the promise of higher statistical precision and improved access to higher $\Delta$m$^2$ phase-space.  
If a resolution of all, rather than most, of the four canonical short-baseline anomalies is desired within the timescale of the next P5 Report, it is essential that experimental efforts along this line be seriously pursued and supported.

In parallel to the efforts described above aimed at directly probing the four anomalies, other currently-running neutrino experiments should continue to provide supporting analyses that constrain specific anomaly interpretations.  
For example, improved IceCube~\cite{IceCube:2020phf,IceCube:2020tka} atmospheric neutrino measurements, NOvA~\cite{NOvA:2017geg,NOvA:2021smv} and T2K~\cite{T2K:2019efw,Bordoni:2017zwi} accelerator decay-in-flight measurements, and KATRIN~\cite{KATRIN:2018sds} and BeEST~\cite{Friedrich:2020nze} beta decay measurements can provide useful new constraints on sterile-mediated flavor mixing scenarios, while the Coherent CAPTAIN-Mills (CCM) experiment~\cite{Aguilar-Arevalo:2021sbh, CCM:2021leg} may provide limits on dark sector phenomena relevant to the LSND or MiniBooNE Anomalies.  
Future high-precision phases of some of these current efforts will also be described in the following section.  

\subsection{Realization of High-Precision Experimental Concepts}

While the experiments mentioned in the previous Section can deliver major advancements in our understanding of the anomalies, some plausible BSM explanations can evade detection in these experiments.  For example, a multi-component heavy neutral lepton sector can deliver a combination of low-amplitude decay and oscillation signatures that may not resolvable by SBN~\cite{Vergani:2021tgc}; in the disappearance sector, reactor-based 3+$N$ $L/E$ oscillation signals may potentially be damped by decoherence effects~\cite{deGouvea:2021uvg,deGouvea:2020hfl}, resulting in untested or ambiguous regions of high-amplitude, high $\Delta m^2$ phase space~\cite{Arguelles:2022bvt}.  
Thus, a robust strategy for resolving the anomalies requires the pursuit of diverse BSM sensitivities well beyond that of existing efforts.  
Thankfully, 
sustained community effort over the past decade has been expended to develop new experimental concepts including cutting-edge sources and/or techniques for probing the primary BSM-related anomaly origin categories---flavor transformation phenomena and new particle production---with significantly greater potential precision and sensitivity.  
Timely installation and completion of these new experimental concepts with different source facilities and detectors would provide a new wave of valuable information well before the close of the coming P5 period, which could be sustained well into the following decade.
The remainder of this section will overview these opportunities for pushing beyond the bounds of the current landscape of anomaly-related results.  
This discussion is roughly organized around the primary BSM anomaly origin focus(es) of each effort: flavor transformation phenomena or new dark sector interactions.

\subsubsection{Efforts Probing Anomalous Flavor Transformation}

To perform highly detailed and systematically independent investigations of the $L/E$- and channel-dependence of short-baseline neutrino flavor transformations, multiple new short-baseline oscillation facilities with varied neutrino sources have been proposed and developed.  Some of these efforts will pursue appearance and/or disappearance measurements with a mix of of muon and electron-flavor fluxes, while others will perform studies of pure-flavor systems.  

\begin{figure}[!ht]
\centering
\includegraphics[width=0.4\textwidth]{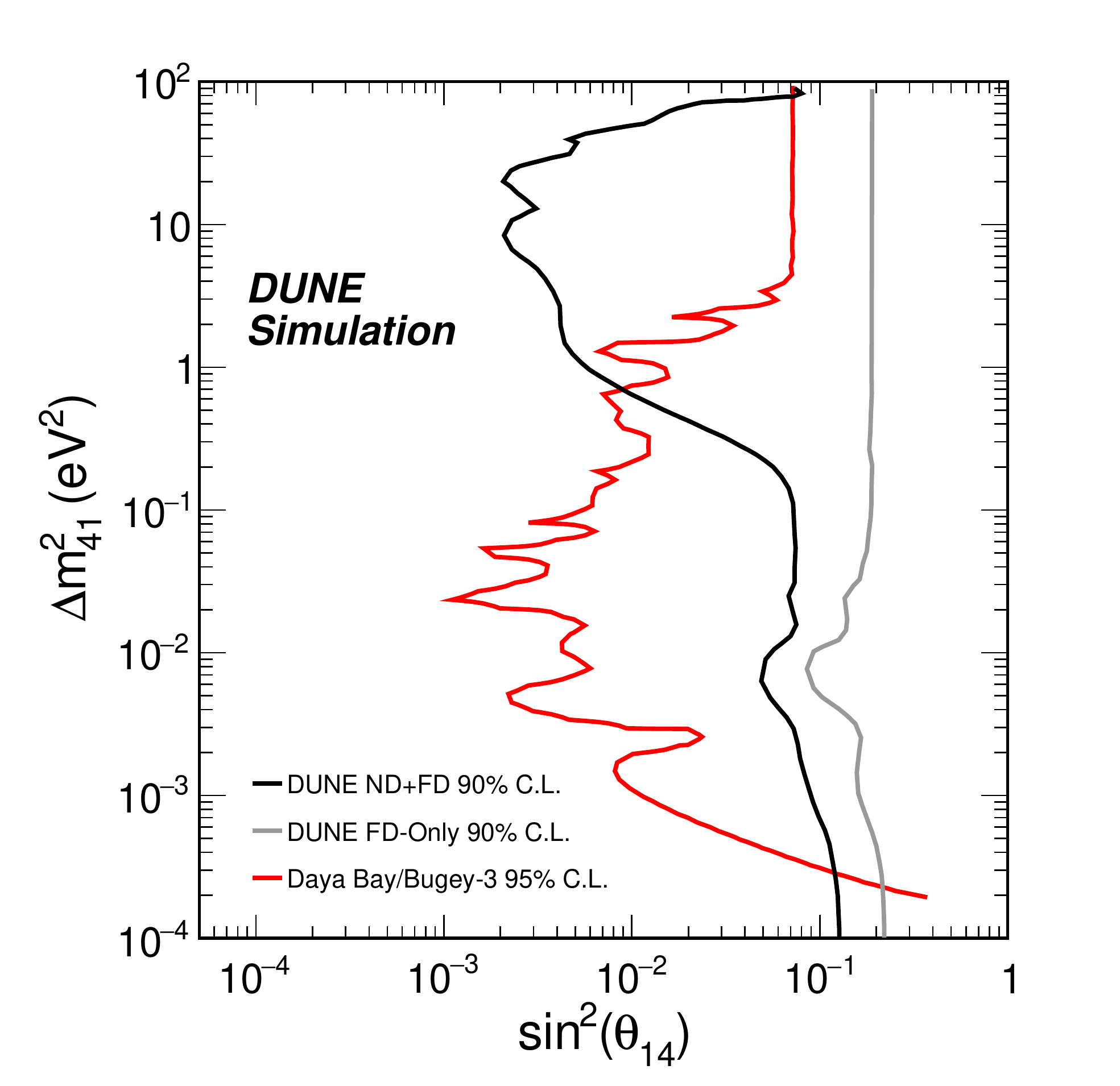}
\includegraphics[width=0.4\textwidth]{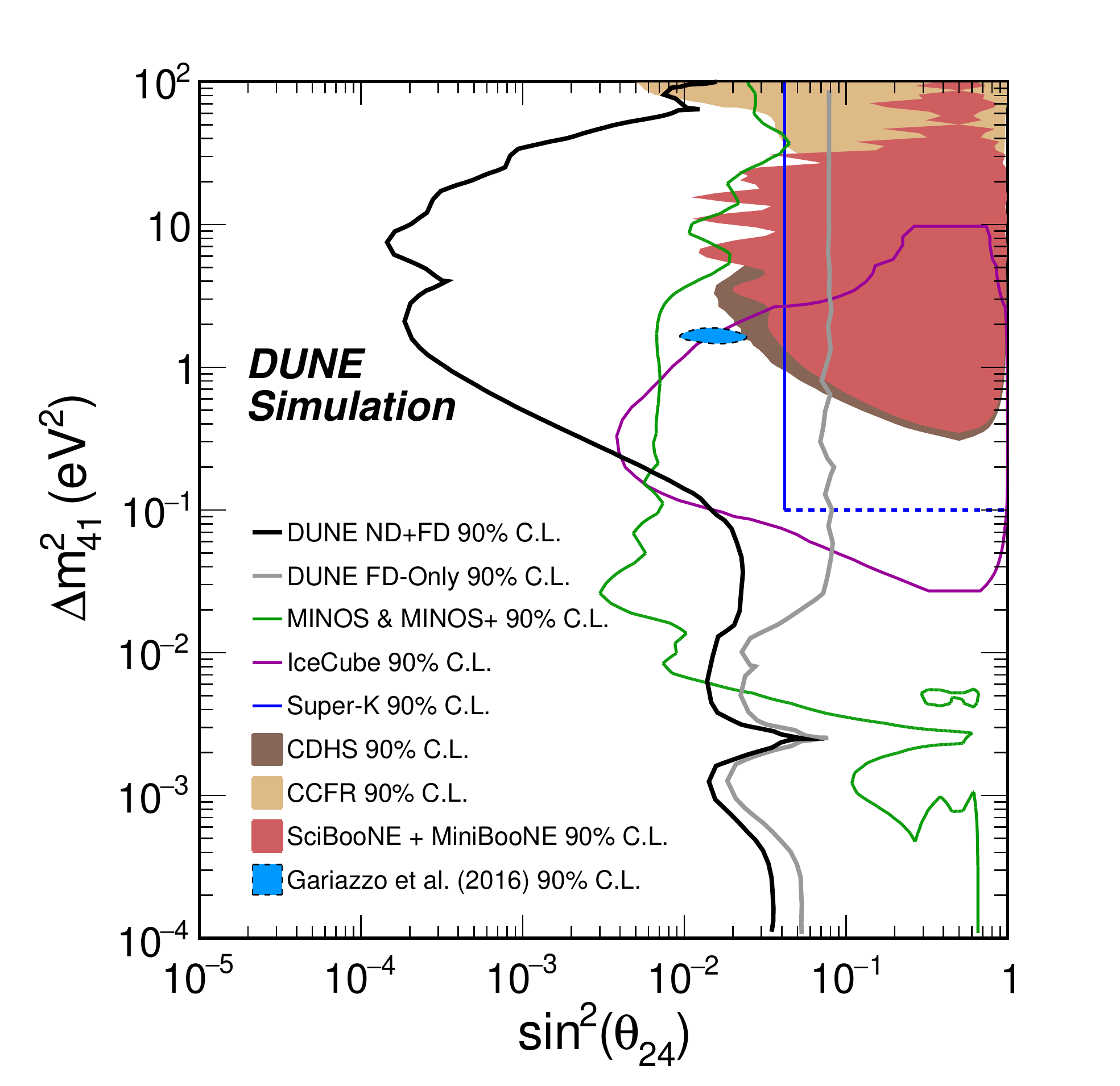}

\caption{Left: DUNE sensitivities to $\theta_{14}$ from the $\nu_e$ CC samples at the ND and FD, assuming $\theta_{24}$=0, along with a comparison with the combined reactor result from Daya Bay and Bugey-3. Right: DUNE sensitivities to $\theta_{24}$ using the $\nu_\mu$ CC and NC samples at both DUNE detectors, along with a comparison with previous and existing experiments.  
}
\label{fig:dune_th_14+th_24}
\end{figure}

The DUNE experiment~\cite{DUNE:2020lwj,DUNE:2020mra,DUNE:2020txw} will serve the future continuation of anomaly-related BSM searches, carrying out short-baseline measurements with its Near Detector (ND) complex, and complementary new physics measurements over the long baseline separating its ND and Far Detector (FD). The DUNE experiment is a next-generation, long-baseline neutrino oscillation experiment, and consists of a high-power, broadband neutrino beam, a powerful precision multi-instrument ND complex located at Fermilab and a 70\,kton LArTPC FD located at the Sanford Underground Research Facility (SURF), 1300\,km away in Lead, South Dakota.  
The wide-band range of energies provided by the LBNF beam affords DUNE significant potential sensitivity to probe sterile mixing, which would typically cause distortions of standard oscillations in energy regions away from the three-flavor $\numu$ disappearance maximum.  
Therefore, DUNE sterile mixing probes reach a broad range of potential sterile neutrino mass splittings by looking for disappearance of charged-current (CC) and neutral-current (NC) interactions over the long distance separating the ND and FD, as well as over the short baseline of the ND.  
DUNE will perform anomalous $\nu_\mu$ and $\nu_e$ disappearance, as well as $\nu_\mu\rightarrow\nu_e$ appearance and NC disappearance oscillation searches as shown in Figure~\ref{fig:dune_th_14+th_24}. 
This figure illustrates the important role a powerful 
ND facility can potentially play in addressing short-baseline oscillation phenomena with DUNE~\cite{DUNE:2020fgq}.  

Other proposed long-baseline neutrino experiments can also provide substantial future BSM-mediated flavor transformation sensitivity relevant to solving the short-baseline anomalies.  
The Hyper-Kamio-kande experiment \cite{Hyper-Kamiokande:2018ofw} plans to perform future sterile neutrino searches with an upgraded ND280 detector and with a new Intermediate Cerenkov Water Detector (IWCD), which can sample multiple neutrino energy ranges with a common moveable water detector.  
A proposed EU-based 2~GeV, 5~MW super neutrino beam facility, ESSnuSB~\cite{ESSnuSB:2013dql, Wildner:2015yaa}, combined with a 500~kT underground water Cherenkov detector, can offer 3+$N$ oscillation sensitivities comparable to those achieved by DUNE.  
A unique search for \numu appearance in a flavor-mixed decay-in-flight \nue$+\bar{\nu}_{\mu}$ beam can be performed using a muon storage ring source, an experimental technique proposed by the nuSTORM experiment~\cite{nuSTORM:2022div}.  
Such an experiment would test the CPT conjugate process of the $\bar{\nu}_{\mu}\rightarrow \anue$ process that defines the most popular explanation for the LSND Anomaly.  
With 10$^{21}$ POT of 60~GeV protons generating $\sim$10$^{18}$ stored muons of 3.8~GeV mean kinetic energy, a 1.3~kT magnetized iron detector at 2~km standoff can generate high confidence level ($>5\sigma$) tests of the best-fit LSND 3+1 oscillation phase space.   

\begin{figure}[!ht]
\centering
\includegraphics[width=0.55\textwidth]{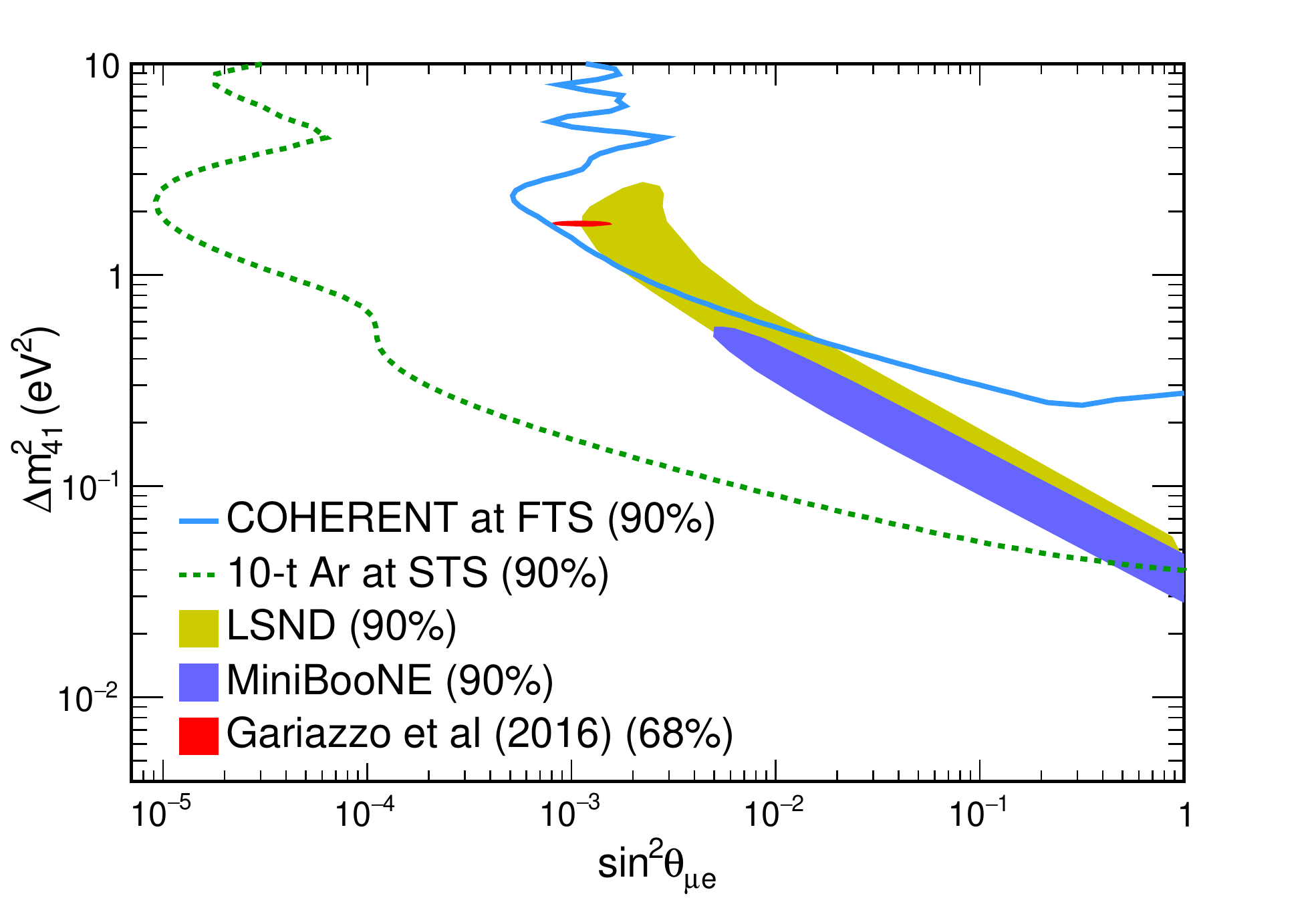}
\includegraphics[width=0.35\textwidth]{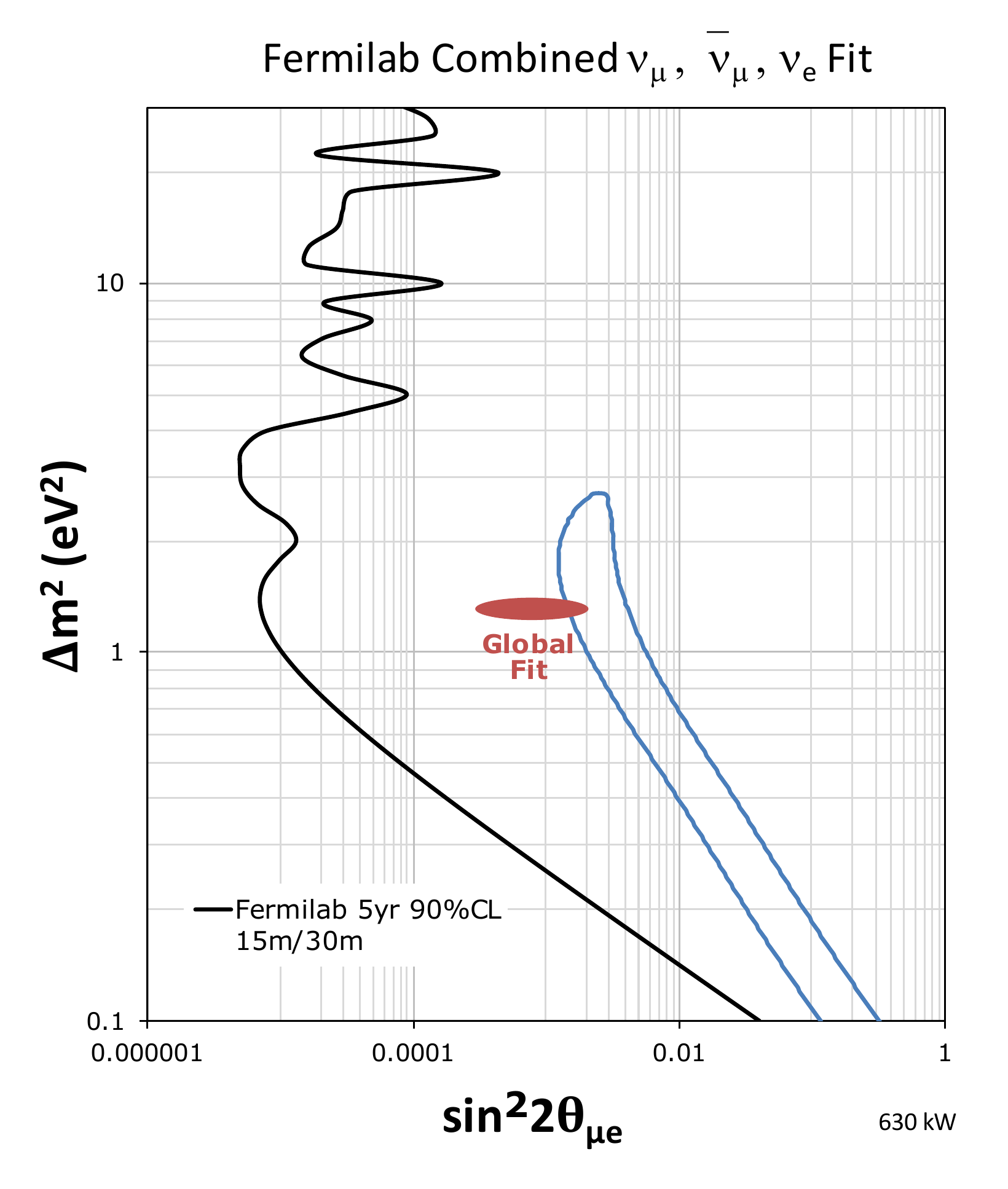}
\caption{Projected sin$^2$2$\theta_{\mu e}$ sensitivity within a 3+1 oscillation framework for future multi-channel decay-at-rest CEvNS measurements by COHERENT and PIP2-BD. Figures taken from Ref.~\cite{Acero:2022wqg}.  
}
\label{fig:dar_cevns}
\end{figure}

Decay-at-rest neutrino sources, such as the the MLF facility used by JSNS$^2$ at J-PARC, the SNS and future STS beamlines at Oak Ridge National Laboratory, and a future PIP-II beam dump facility at Fermilab~\cite{Toups:2022knq,Toups:2022yxs}, represent promising mixed-flavor sources that offer the potential for BSM flavor transformation studies capable of pushing beyond the projected $\bar{\nu}_{\mu}$ $\rightarrow$ \anue reach of JSNS$^2$/JSNS$^2$-II.  
The short beam width of and time-ordered decay of 29.8~MeV \numu-producing stopped pions and $\mathcal{O}(0-50)$~MeV~$\anumu$- and $\nue$-producing stopped muons in these sources generates neutrino emissions with a well-defined, time-dependent flavor structure.  
This time structure, when coupled with a short-baseline detector sensitive to NC interactions, such as low-threshold coherent elastic neutrino-nucleus scattering (CEvNS) detectors, can enable simultaneous measurement of both electron-flavor and muon-flavor disappearance.  
Within a 3+1 framework, combination of these measurements can yield wide sensitivity to parameters governing $\numu$ $\rightarrow$ \nue oscillation~\cite{Acero:2022wqg}.  
At Oak Ridge National Laboratory, CEvNS-based oscillation searches have been proposed by the COHERENT collaboration using near-phase large detector deployments within Neutrino Alley, which sits at roughly the first oscillation maximum for $\Delta$m$^2\sim1$~eV$^2$.  
On longer timescales, such measurements could be performed in a 10-ton-scale detector in a dedicated neutrino facility to be built in the next decade between the targets of the SNS and future STS beamlines.  
Example sensitivity of proposed near-term (long-term) ORNL-based CEvNS measurements of 3~years (5~years) duration to sin$^2$2$\theta_{\mu e}$ within a 3+1 oscillation framework are shown in Figure~\ref{fig:dar_cevns}.  
At Fermilab, the 800~MeV PIP-II LINAC feeding the LBNF neutrino beamline could also be coupled to a new Booster-sized permanent magnet or DC-powered accumulator ring and possibly a new rapid-cycling synchotron, yielding a high-powered proton beam dump neutrino source~\cite{Toups:2022yxs}.  
The 5 year sin$^2$2$\theta_{\mu e}$ sensitivity of PIP2-BD, consisting of a 630~kW, 1.2~GeV accumulator paired with two 100~t LAr scintillation CEvNS detectors at 15~m and 30~m distances, is also shown in Figure~\ref{fig:dar_cevns}.

\begin{figure}[htbp!]
    \centering
    \includegraphics[width=0.4 \textwidth]{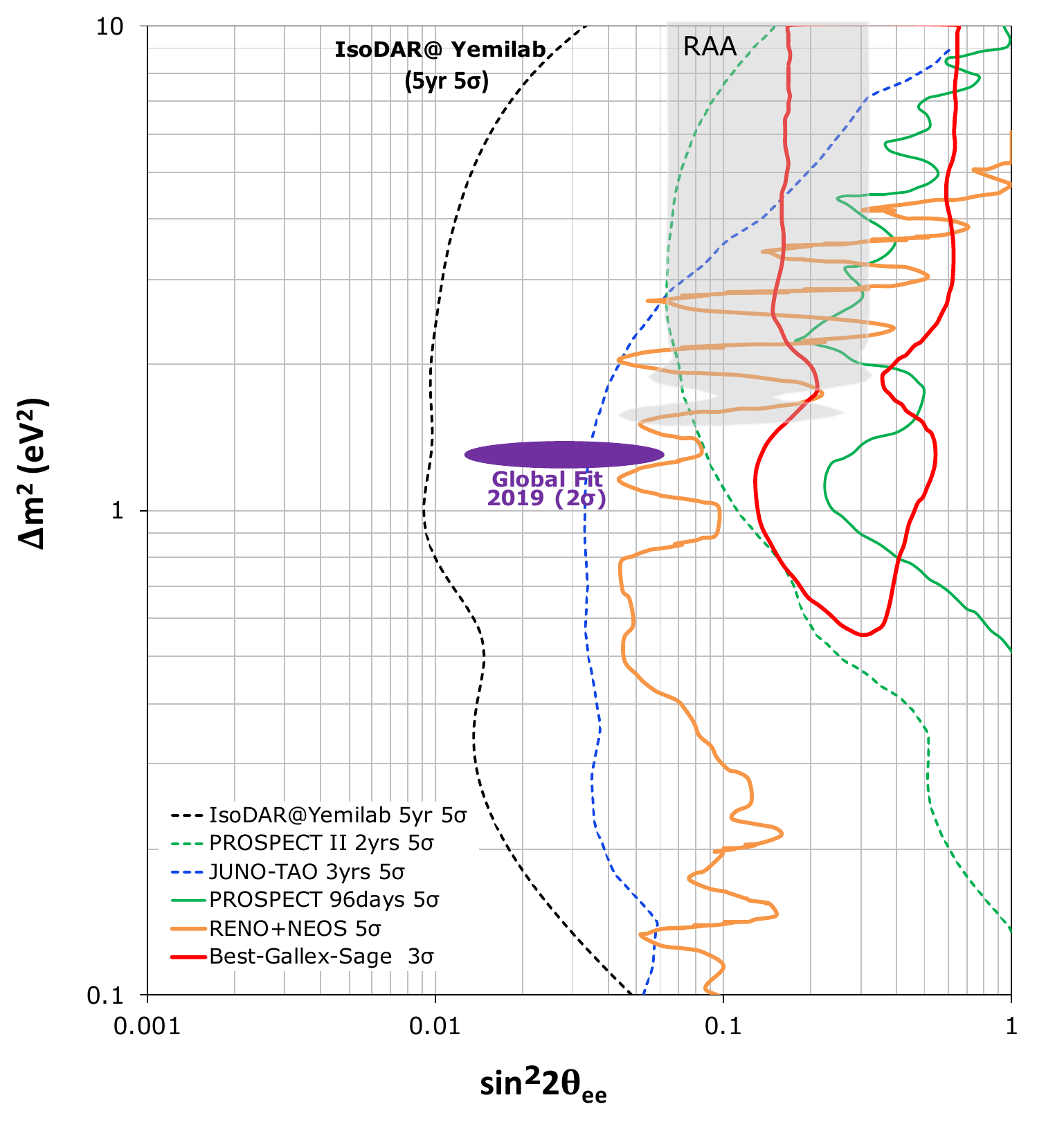}
    \includegraphics[width=0.58 \textwidth]{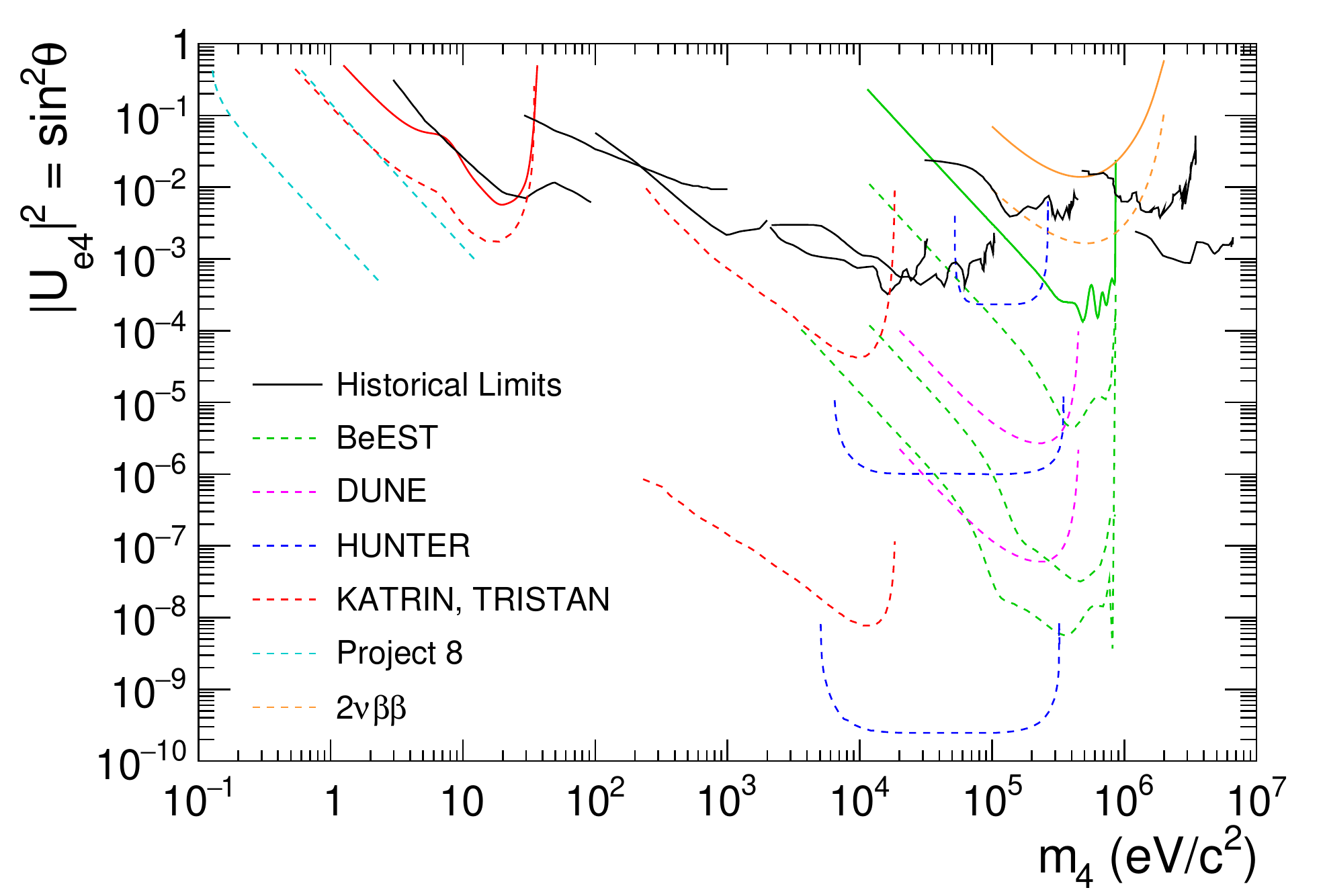}
    \caption{
        Left: 3+1 \anue disappearance oscillation sensitivity projections for the 5~year Yemilab-based IsoDAR experiment described in the text; also pictured are exclusion (sensitivity) curves for NEOS+RENO and PROSPECT (JUNO-TAO and PROSPECT), and allowed regions for the Gallium and Reactor Anomalies.  
        Right: Existing exclusions (solid) and projected sensitivities (dotted) for active-sterile couplings from measurements of weak nuclear decay products.  
        Note that similar phase space regions are being pictured, but with axes reversed, and with mass axes either squared (left) or not (right).  
    } 
    \label{fig:nuedis}
  \end{figure}

Future high-precision searches for anomalous neutrino flavor transformation should also be pursued using flavor-pure neutrino sources.  
In the stable of recently-completed experimental efforts, sterile-mediated electron flavor disappearance has been probed with high purity using short-baseline reactor \anue experiments at MeV scales.  
In the near future, electron-flavor transitions could be probed with substantially higher precision than projected by current and near-future short-baseline reactor experiments using a cyclotron-based $^8$Li isotope decay-at-rest (IsoDAR~\cite{Alonso:2022mup}) neutrino source deployed near a large underground detector.   
When the IsoDAR neutrino source with $\sim$3-13~MeV \anue energies is paired with a 2.3 kton Yemilab-based liquid scintillator detector (LSC), approximately 1.6 million inverse beta decay events are expected to be reconstructed in 5 years of running~\cite{Seo:2019dpr}.  
The high statistics and large energy and baseline spread enable strong capabilities for precisely studying the $L/E$ dependence of short-baseline disappearance and differentiating between potential BSM flavor transformation models, such as 3+1 oscillations, 3+1+decay, and more~\cite{Alonso:2021kyu}.  
The 3+1 oscillation sensitivity of the Yemilab arrangement described above is pictured in Figure~\ref{fig:nuedis}; this experiment is projected to be capable of probing at 5$\sigma$ confidence level \anue disappearance amplitudes to the few-percent level across all $\Delta$m$^2$ from 0.1-10~eV$^2$.  
IsoDAR has received preliminary approval to run at Yemilab in the configuration described above, and Yemilab caverns capable of hosting these facilities have been excavated.  

Flavor-pure searches for \numu disappearance have been proposed using 235.5~MeV kaon decay-at-rest neutrinos produced by the future PIP-II beam dump facility described earlier in this section.  This source, when combined with an elongated KPIPE-style scintillator detector~\cite{Axani:2015dha}, could generate a \numu disappearance dataset rivalling or exceeding the muon-flavor disappearance capabilities of the SBN program, particularly at higher ($\gtrsim10$~eV$^2$) $\Delta$m$^2$.

Intrinsically flavor-pure beta decay and electron capture experiments are complementary to oscillation-based searches for electron-flavor couplings to new neutrino mass states.  
By exploiting the kinematics of the weak process, these experiments have the capability to search for states ranging in mass from the eV to 100's of keV~scale.  
The presence of new states will be observed in these experiments via observations of distortions of a theoretically-predicted beta decay spectrum, or through detection of missing energy among all final-state decay products.  
The mass range addressed by each of the various experiments is primarily dictated by the endpoint of the experiment's radioactive isotopes and the experiment's achievable energy resolution.  
Many of these current and proposed future experiments have the potential to offer strong sensitivity to small active-sterile couplings, either due to the use of intense radioactive sources or extremely large detector masses.  
The former category includes KATRIN/TRISTAN~\cite{KATRIN:2018oow,Mertens:2020mdv}, Project-8~\cite{Project8:2017nal}, HUNTER~\cite{Martoff:2021vxp}, and BeEST~\cite{Leach:2021bvh}, while the latter category includes DUNE (far detector)~\cite{WARP:2006nsa, DUNE:2020lwj, DUNE:2020mra, DUNE:2020txw} and next-generation 0$\nu\beta\beta$ experiments.  
Projected sensitivities for these efforts are presented in Figure~\ref{fig:nuedis}.  
Potential contributions to probing this parameter space are also possible from future solar neutrino measurements to be carried out by large dark matter experiments or by large-volume far detectors at long-baseline experiments, as pointed out in Ref.~\cite{Goldhagen:2021kxe}.

Beyond the BEST experiment follow-ups described in the previous section, some longer-term experimental proposals aim to use intense $^{51}$Cr or $^{144}$Ce sources at large underground scintillator detectors to perform \anue disappearance studies while directly confronting the Gallium Anomaly.  
In China, the Jinping solar neutrino experiment has studied deployment of a 100~kCi $^{144}$Ce-$^{144}$Pr source either in the center or just outside of its 2~kT scintillator target \cite{Smirnov:2020bcr}, similar to that attempted by the cancelled SOX experimental effort.  
The THEIA collaboration is also investigating the benefits of deploying either $^{51}\rm{Cr}$ or $^{144}$Ce-$^{144}$Pr sources inside a future very large (>20~kt) detector composed of either liquid scintillator or water-based liquid scintillator \cite{OrebiGann:2015gus, Theia:2019non}.   

Finally, atmospheric neutrino experiments will continue to offer unique insight into flavor transformation interpretations of the anomalies by offering unique sensitivity to the tau content of new neutrino mass states, as well as access to resonant enhancements associated with that state or other effects associated with neutrino propagation, e.g.~neutrino decay.  
Current experiments, such as Super-Kamiokande and IceCube, have recently provided intriguing demonstrations of these capabilities, in some cases providing sterile oscillation exclusion power rivalling or exceeding those of comparable accelerator-based measurements \cite{IceCube:2017ivd}. 
The impending or proposed efforts of DUNE, Hyper-Kamiokande, IceCube Upgrade \cite{IceCube:2019pna}, ORCA \cite{KM3NeT:2021ozk}, and THEIA \cite{Theia:2019non} offer sensitivities exceeding those of these two mentioned predecessors due to either their larger target size or their enhanced energy threshold or calibration techniques.  
 
 \subsubsection{Efforts Probing Dark Sector Interactions}
 
 Whether the anomalies are caused by a unified but complex regime of new physics, or by individual systematic or BSM effects specific to subsets of neutrino source types, the adoption of a multi-probe experimental test strategy is essential to their convincing resolution. 
 Specifically, exciting new physics scenarios developed through joint efforts by the theoretical and experimental communities posit that the MiniBooNE Anomaly may derive from couplings to a new hidden sector. 
 This possibility can be probed within an expansive, global new physics research program taking full advantage of the world's foremost proton and electron beam facilities, to be developed in close collaboration with the NF03, RF06, and other Frontier Sub-Topic communities.  
 
\begin{figure}[ht!]
    \centering
    \includegraphics[width=0.32 \textwidth]{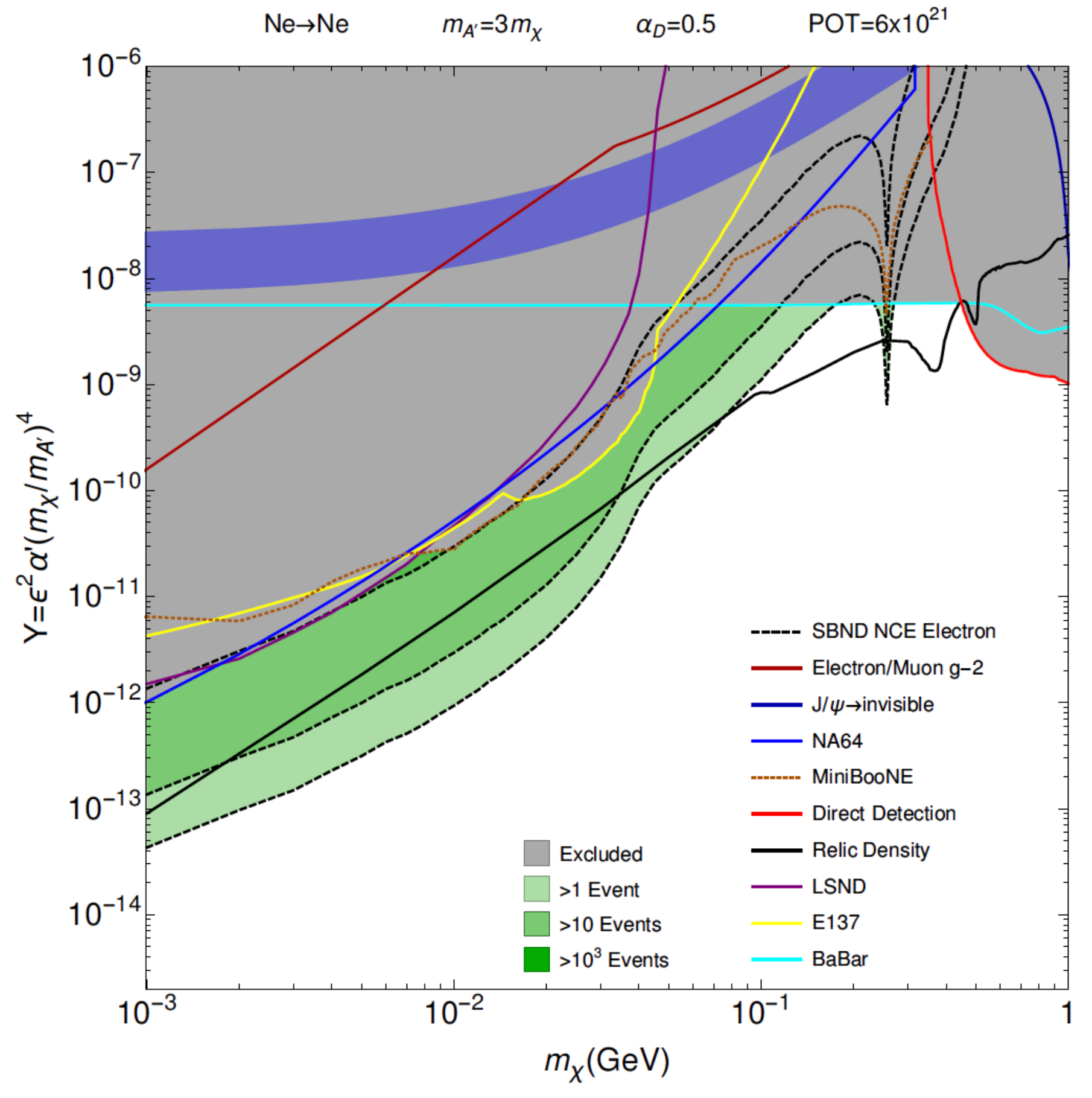}
    \includegraphics[width=0.33 \textwidth]{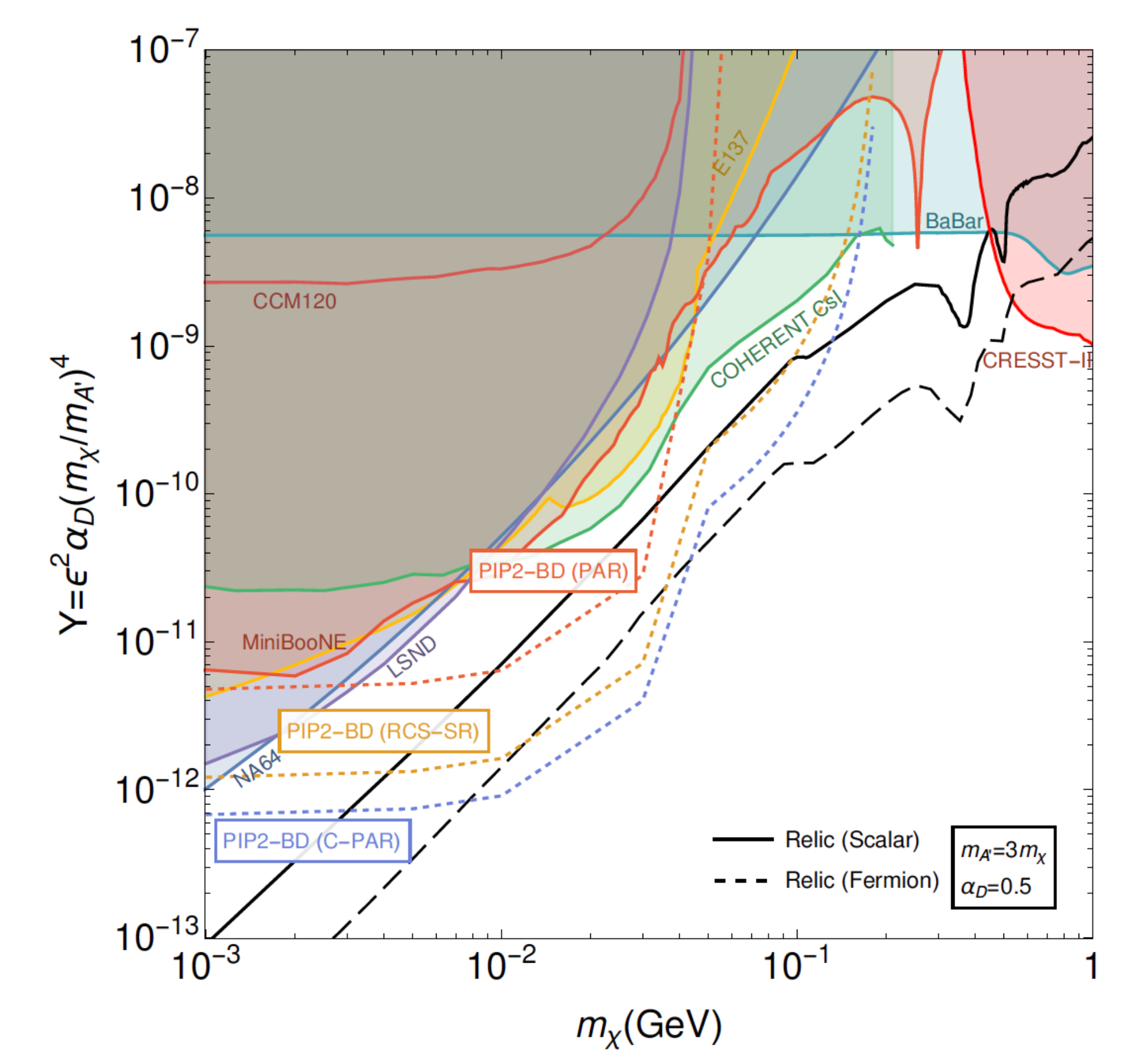}
    \includegraphics[width=0.32 \textwidth]{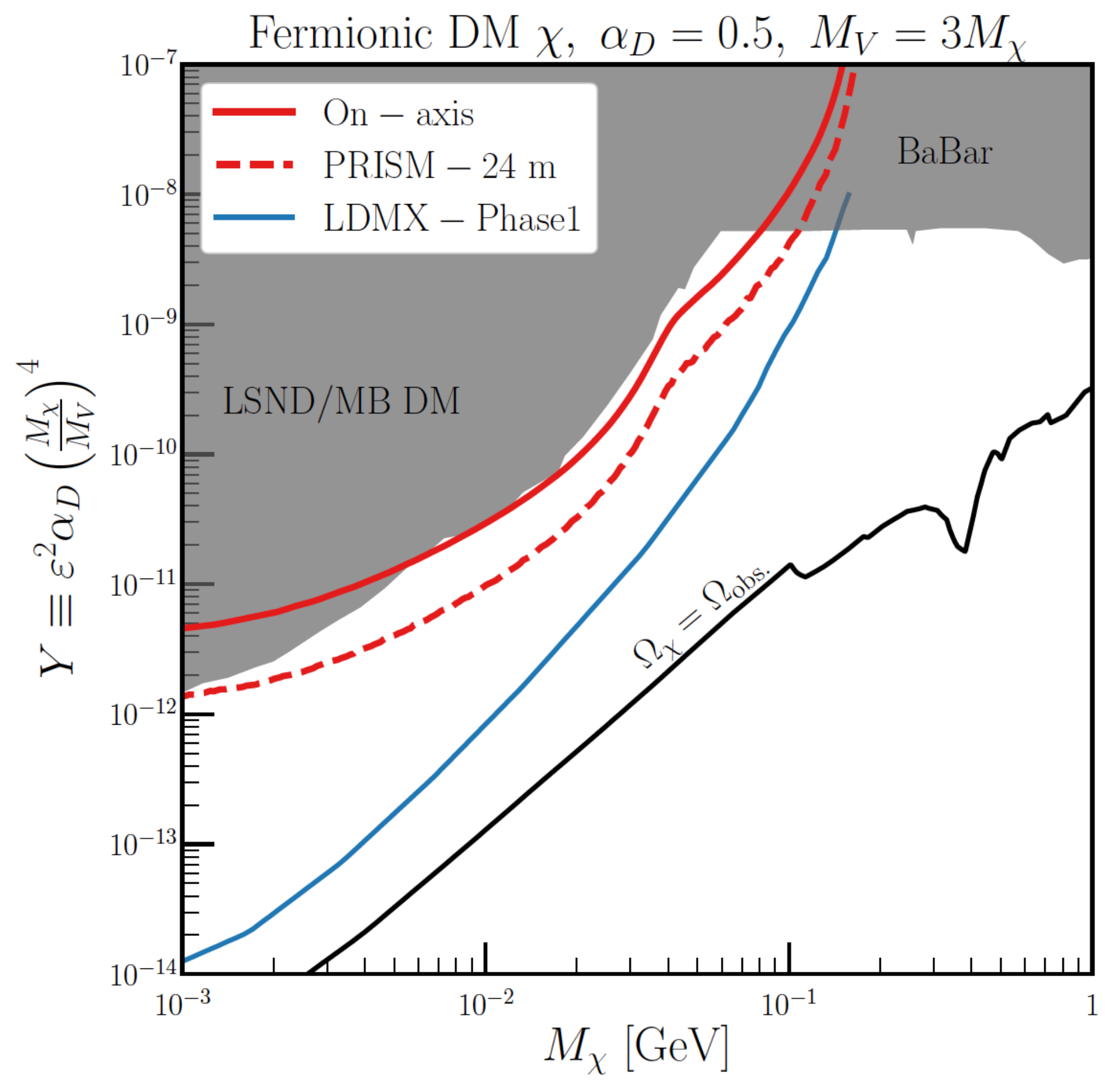}
    \caption{
        Regions of relic abundance parameter $Y$ vs. dark matter mass $m_\chi$. The solid black line is the scalar relic density line that can be probed.
        Left: Signal sensitivity expected for SBN-BD using the electron scattering channel. 
        Center: Signal sensitivity expected for PIP2-BD using the CE$\nu$NS channel Right: Signal sensitivity expected for the Phase I DUNE ND using the electron scattering channel. The dashed red line shows possible improvements from using the PRISM concept. Further improvements would be possible with the full-scope DUNE ND complex.  
    } 
    \label{fig:sbn_pip2_dune_dm}
  \end{figure}

Significant opportunities in this direction are offered by liquid argon time projection chambers~\cite{Machado:2019oxb} due to the capabilities of these detectors in discriminating different signatures, such as $e^-, \, e^+e^-, \,  \gamma,$ and $\gamma\gamma$ final states, as well as to identify final state hadron multiplicities. 
These LArTPC capabilities would allow for vertex identification, which in turn allows for searching for displaced decay positions. 
In particular,
the possibility of antineutrino running or beam dump running in SBN~\cite{Toups:2022knq} (which would require BNB lifetime extension), and the deployment of the full-scope DUNE-ND complex~\cite{DUNE:2020fgq}, would provide invaluable opportunities to probe these scenarios. 
In addition, deployment of a 100 ton LAr scintillation-only detector downstream of the PIP2 beam dump~\cite{Toups:2022yxs} could provide leading reach for these probes.
Expected sensitivities achievable for these efforts are shown in Figure~\ref{fig:sbn_pip2_dune_dm}.
These scenarios could also be probed at Hyper-Kamiokande's near detector, and the near detector envisioned for nuSTORM~\cite{nuSTORM:2022div} would enable sensitivity to models with neutrino upscattering to new dark particles with decays to $e^+e^-$.

Additionally, higher energy experiments can also contribute to searches for these new particles.
Atmospheric neutrinos could be used to probe dark sectors in which neutrino interactions, either in the detector or in the surrounding dirt, lead to the production of new particles. 
The signals of the decay or scattering of these particles could be fairly different from the atmospheric neutrino background, and allow for leveraging the high statistical sample in large neutrino detectors.
For example, double-bang signatures similar to tau neutrino scatterings at IceCube and its upgrade~\cite{IceCube:2019pna} would be a signature of heavy neutral leptons and dark neutrinos.
Similar arguments would apply to new physics searches in KM3NET~\cite{Adrian-Martinez:2016fdl} and ORCA~\cite{KM3NeT:2021ozk}, as well as high-precision and smaller volume detectors such as DUNE~\cite{DUNE:2020ypp}, Hyper-Kamiokande~\cite{Hyper-Kamiokande:2018ofw}, JUNO~\cite{JUNO:2020ijm} and THEIA~\cite{Theia:2019non}.
High energy accelerator neutrino experiments such as FASER$\nu$ \cite{FASER:2019dxq} and SND@LHC \cite{SHiP:2020sos}, as well as proposed experiments which go under the umbrella term of Forward Physics Facility (FPF), would be also sensitive to signatures of relatively heavier dark sectors due to their neutrino energy spectra at the 100~GeV--TeV window.
 
\begin{figure}[ht!]
    \centering
    \includegraphics[width=0.76 \textwidth]{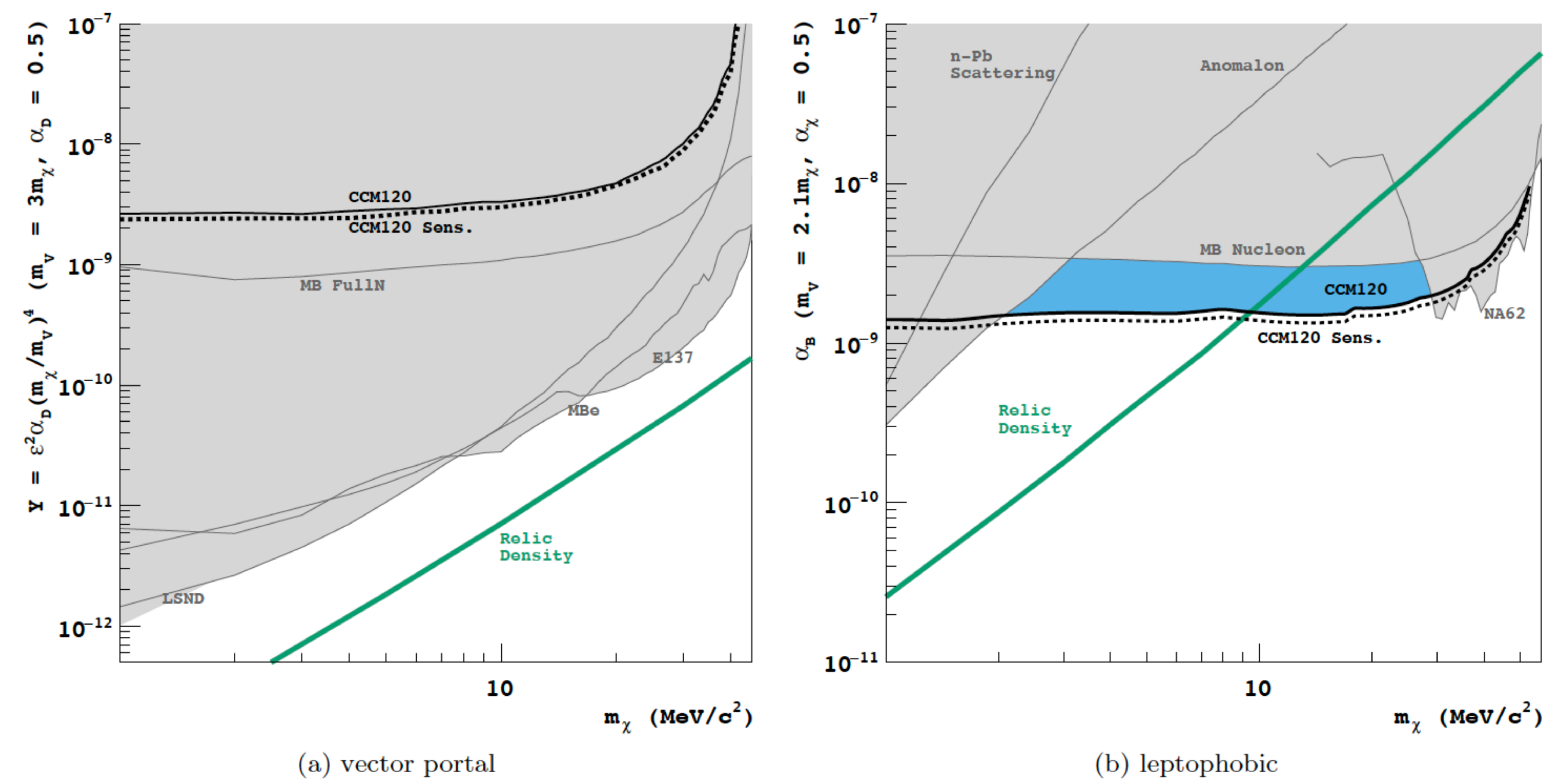} 
    \caption{
        The median sensitivity (dashed) and 90\% C.L. (solid) results from CCM120 compared to existing limits for (a) vector portal model; and (b) leptophobic model.
    } 
    \label{fig:CCM_dm}
  \end{figure}

Further, the CCM experiment at LANL \cite{osti_1572310} is using the LANSCE beamline and a 10~ton liquid argon detector to probe a wide variety of anomaly-relevant hidden sector particles and forces, including dark photons, axion-like particles, and heavy neutral leptons in the keV to MeV mass range~\cite{Dutta:2021cip}.
Many of these particles are invoked as alternative or additional explanations to oscillations involving sterile neutrinos as the source of MiniBooNE Anomaly. As an example, the first results from CCM120~\cite{CCM:2021leg}, which uses 120 PMTs for detector coverage, are shown in Figure~\ref{fig:CCM_dm}. Significant improvements are expected after deployment of the full complement of 200 PMTs in CCM200. 

Other infrastructure used primarily for non-neutrino physics purposes, such as CERN's kaon decay-in-flight facility, host of the NA62 experiment, can also expand experimental coverage of anomaly-relevant BSM phase-space \cite{Anchordoqui:2021ghd}.  
This facility enables searches for production of new heavy (up to hundreds of MeV) neutrino mass states or associated mediators.


\section{Further Guiding Principles for Next Decade's Efforts}

As the short-baseline anomalies are explored and better understood over the next decade, the community's efforts should be directed toward disentangling the plethora of possibilities that have been identified as viable interpretations, following the opportunities outlined in the previous Section.  
Beyond the priorities of supporting existing and future experimental and theoretical efforts, we highlight five broad conceptual themes that should also be reflected in the next decade's efforts.

\begin{itemize}
\item{\textbf{Cover all anomaly sectors:} Given the unresolved nature of the anomalies and their interplay under different interpretations, it is imperative to support all pillars of a diverse experimental portfolio---source, reactor, decay-at-rest, decay-in-flight, and other methods/sources---to provide complementary probes of and increased precision for new physics explanations.
}
\item{\textbf{Pursue diverse signatures:} Given the diversity of possible experimental signatures associated with allowed anomaly interpretations, it is imperative that current and future experiments make design and/or analysis choices that maximize sensitivity to as broad an array of these potential signals as possible.
}
\item{\textbf{Strengthen theoretical engagement:} Priority in the theory community should be placed on the development of new physics models collectively relevant to all four canonical short-baseline anomalies and the development of tools for enabling efficient tests of these models with existing and future experimental datasets.
}
\item{\textbf{Openly share data:} Fluid communication between the experimental and theory communities will be required, which implies that both experimental data releases and theoretical calculations, and global fitting tools should be publicly available.  
Data and tool sharing is particularly important, as it is likely that resolution of the anomalies will require a combination of measurements.
}

\item{\textbf{Apply robust analysis techniques:} Appropriate statistical treatment is crucial to quantify the compatibility of data sets within the context of any given model, and in order to test the absolute viability of a given model. 
Accurate evaluation of allowed parameter space is also an important input to the design of future experiments.
}
\end{itemize}



\section{Conclusion}

Full exploitation of recent and ongoing experimental and theory efforts provides valuable new information and plausible paths to convergence for interpretation of the short-baseline anomalies, or potentially to exciting new physics discoveries. These developments and future advancements are a priority within the community, and it is likely that they will cause substantial shifts in the short-baseline experimental and theoretical landscape, potentially as early as by the middle of the upcoming P5 period, with subsequent priorities depending on what is learned about the various hypothesized anomaly origins.  
To prepare for this likelihood, the community should work presently to explore, develop, and support diverse opportunities for new dedicated measurements at existing or future facilities, and it should also be prepared to re-evaluate priorities given what will be learned. 
Beyond the community, it is also essential that available supporting resources are structured to enable and facilitate adaptation to developments during the next P5 period, and to ensure that timely experimental groundwork is laid for following generations of precision short-baseline measurements.


\cleardoublepage

\section*{Appendix: LOIs submitted to NF02}
\markboth{Appendix: LOIs submitted to NF02}{}

There were 47 Letters of Interest (LOIs) submitted to the NF02 group. Links to each letter listed in Table~\ref{tab:loi} are found on the web page:
\url{https://snowmass21.org/neutrino/sterile/start}

\begin{table}[!htp]\centering
\scalebox{0.7}{
\scriptsize
\tabcolsep=0.11cm
\begin{tabular}{clll}\toprule
Index &\cellcolor[HTML]{e2efda}Filename &\cellcolor[HTML]{e2efda}Title \\\cmidrule{1-3}
1 &SNOWMASS21-AF5\_AF0-NF2\_NF0-RF6\_RF0\_Vandewater-215.pdf &LANSCE-PSR Short-Pulse Upgrade for Improved Dark Matter and Sterile Neutrino Searches \\
2 &\cellcolor[HTML]{e2efda}SNOWMASS21-CF1\_CF0-NF3\_NF2-IF1\_IF2\_Hubbard-203.pdf &\cellcolor[HTML]{e2efda}Dark Matter Searches with the Micro-X X-ray Sounding Rocket \\
3 &SNOWMASS21-CF1\_CF3-NF2\_NF3-TF8\_TF9\_Bibhushan\_Shakya-267.pdf &Indirect Detection Aspects of Hidden Sector Dark \\
4 &\cellcolor[HTML]{e2efda}SNOWMASS21-CF5\_CF3-NF2\_NF0-TF9\_TF11\_Neelima\_Sehgal-016.pdf &\cellcolor[HTML]{e2efda}CMB-HD An Ultra-Deep, High-Resolution \\
5 &SNOWMASS21-CF7\_CF3-NF3\_NF2-TF9\_TF0\_Joel\_Meyers-147.pdf &Insights for Fundamental Physics and Cosmology with Light Relics \\
6 &\cellcolor[HTML]{e2efda}SNOWMASS21-EF7\_EF0-NF2\_NF3-RF4\_RF0\_Brian\_Batell-114.pdf &\cellcolor[HTML]{e2efda}BSM Physics at the Electron Ion Collider Searching for Heavy Neutral Leptons \\
7 &SNOWMASS21-IF3\_IF8-NF2\_NF9\_Jing\_Liu-095.pdf &COHERENT 5 Instrumentation Development \\
8 &\cellcolor[HTML]{e2efda}SNOWMASS21-NF1\_NF2\_Daya\_Bay-086.pdf &\cellcolor[HTML]{e2efda}Legacy of the Daya Bay Reactor Neutrino Experiment \\
9 &SNOWMASS21-NF2\_NF0-CF1\_CF0\_Peter\_Meyers-018.pdf &HUNTER A Facility for a Trapped Atom Sterile \\
10 &\cellcolor[HTML]{e2efda}SNOWMASS21-NF2\_NF0-CF7\_CF0-TF9\_TF11-200.pdf &\cellcolor[HTML]{e2efda}Sterile neutrinos with non-standard interactions \\
11 &SNOWMASS21-NF2\_NF0-IF3\_IF0\_Susanne\_Mertens-197.pdf &Prospects for keV Sterile Neutrino Searches with KATRIN \\
12 &\cellcolor[HTML]{e2efda}SNOWMASS21-NF2\_NF0\_SBN-164.pdf &\cellcolor[HTML]{e2efda}Sensitive Tests for Sterile Neutrino Oscillations at the Short-Baseline Neutrino Program at Fermilab \\
13 &SNOWMASS21-NF2\_NF1\_Joint\_Oscillation\_Analyses\_at\_Reactors-115.pdf &Joint Experimental Oscillation Analyses in Search of Sterile Neutrinos \\
14 &\cellcolor[HTML]{e2efda}SNOWMASS21-NF2\_NF1\_Rosner-045.pdf &\cellcolor[HTML]{e2efda}THREE STERILE NEUTRINOS IN E6 \\
15 &SNOWMASS21-NF2\_NF3-EF9\_EF0-RF4\_RF6-CF1\_CF0-TF8\_TF11\_Matheus\_Hostert-041.pdf &Opportunities and signatures of non-minimal Heavy \\
16 &\cellcolor[HTML]{e2efda}SNOWMASS21-NF2\_NF3-RF6\_RF0\_Athanasios\_Hatzikoutelis-160.pdf &\cellcolor[HTML]{e2efda}Physics Opportunities for detection and study of Heavy \\
17 &SNOWMASS21-NF2\_NF3-TF11\_TF8\_D-V-Forero-078.pdf &Large Extra-Dimension Searches \\
18 &\cellcolor[HTML]{e2efda}SNOWMASS21-NF2\_NF3\_Alex\_Sousa-150.pdf &\cellcolor[HTML]{e2efda}Long-Baseline Accelerator Probes for Light Sterile2Neutrinos \\
19 &SNOWMASS21-NF2\_NF3\_Arindam\_Das-036.pdf &NSI from a ﬂavorful Z(cid48) model \\
20 &\cellcolor[HTML]{e2efda}SNOWMASS21-NF2\_NF3\_G\_V\_Stenico-131.pdf &\cellcolor[HTML]{e2efda}Neutrino Decay as a Solution to the Short-Baseline Anomalies \\
21 &SNOWMASS21-NF2\_NF3\_Gavin\_Davies-117.pdf &The NOvA Experiment and Exotic Neutrino Oscillations \\
22 &\cellcolor[HTML]{e2efda}SNOWMASS21-NF2\_NF3\_Michael\_Mooney-196.pdf &\cellcolor[HTML]{e2efda}Search for Heavy Sterile Neutrinos Using 39Ar Beta \\
23 &SNOWMASS21-NF2\_NF3\_carlo.giunti@to.infn.it-008.pdf &NuSte Global Light Sterile Neutrino Fits \\
24 &\cellcolor[HTML]{e2efda}SNOWMASS21-NF2\_NF4\_Sousa\_Thakore-134.pdf &\cellcolor[HTML]{e2efda}Sterile Neutrino Searches with Atmospheric Neutrinos \\
25 &SNOWMASS21-NF2\_NF6-CF1\_CF0-IF8\_IF0\_Bob\_Wilson-079.pdf &ICARUS in the Next Decade \\
26 &\cellcolor[HTML]{e2efda}SNOWMASS21-NF2\_NF7\_Dazeley-149.pdf &\cellcolor[HTML]{e2efda}An Application of Pulse Shape Sensitive Plastic Scintillator - Segmented AntiNeutrino \\
27 &SNOWMASS21-NF2\_NF7\_Jon\_Link-075.pdf &CHANDLER A Technology for Surface-level Reactor Neutrino Detection \\
28 &\cellcolor[HTML]{e2efda}SNOWMASS21-NF2\_NF9-080.pdf &\cellcolor[HTML]{e2efda}Neutrino Physics with IsoDAR \\
29 &SNOWMASS21-NF2\_NF9-CF1\_CF0\_Jon\_Link-038.pdf &Physics with Electron Capture Neutrino Sources \\
30 &\cellcolor[HTML]{e2efda}SNOWMASS21-NF2\_NF9\_Further\_PROSPECT-I\_Science-168.pdf &\cellcolor[HTML]{e2efda}Forthcoming Science from the PROSPECT-I Data Set \\
31 &SNOWMASS21-NF2\_NF9\_Future\_PROSPECT-II\_Science-169.pdf &The Expanded Physics Reach of PROSPECT-II \\
32 &\cellcolor[HTML]{e2efda}SNOWMASS21-NF2\_NF9\_Spitz-128.pdf &\cellcolor[HTML]{e2efda}The JSNS2 Experiment \\
33 &SNOWMASS21-NF3\_NF2-CF7\_CF0\_Parno-042.pdf &Searches for Beyond-Standard-Model Physics with the KATRIN Experiment \\
34 &\cellcolor[HTML]{e2efda}SNOWMASS21-NF3\_NF2-CF7\_CF1-TF9\_TF8\_Katori-073.pdf &\cellcolor[HTML]{e2efda}New physics with astrophysical neutrino ﬂavor \\
35 &SNOWMASS21-NF3\_NF2-RF3\_RF0-IF1\_IF0\_Kyle\_Leach-055.pdf &Laboratory-Based keV-Scale Sterile Neutrino Searches \\
36 &\cellcolor[HTML]{e2efda}SNOWMASS21-NF3\_NF2-TF11\_TF0-019.pdf &\cellcolor[HTML]{e2efda}Testing quasi-Dirac leptogenesis through neutrino oscillations \\
37 &SNOWMASS21-NF3\_NF2-TF11\_TF0\_DUNE-051.pdf &Physics Beyond the Standard Model in DUNE \\
38 &\cellcolor[HTML]{e2efda}SNOWMASS21-NF3\_NF2-TF11\_TF0\_Petrillo-189.pdf &\cellcolor[HTML]{e2efda}Follow up of anomalies measured in short baseline \\
39 &SNOWMASS21-NF3\_NF2\_Ben\_Jones-046.pdf &BSM Neutrino Oscillation Searches with 1-100 TeV \\
40 &\cellcolor[HTML]{e2efda}SNOWMASS21-NF3\_NF2\_Celio\_Moura-063.pdf &\cellcolor[HTML]{e2efda}Non-Unitarity of the neutrino mixing matrix \\
41 &SNOWMASS21-NF5\_NF2\_Project8\_Secondary\_Physics-171.pdf &Secondary Physics Potential of the Project 8 Experiment \\
42 &\cellcolor[HTML]{e2efda}SNOWMASS21-NF6\_NF2-EF0\_EF0-AF2\_AF4-077.pdf &\cellcolor[HTML]{e2efda}nuSTORM collaboration \\
43 &SNOWMASS21-NF6\_NF2-EF0\_EF0-AF2\_AF4\_Kenneth\_Long-082.pdf &nuSTORM collaboration \\
44 &\cellcolor[HTML]{e2efda}SNOWMASS21-NF6\_NF2\_Kenneth\_Long-076.pdf &\cellcolor[HTML]{e2efda}nuSTORM collaboration \\
45 &SNOWMASS21-RF4\_RF0-NF3\_NF2-TF11\_TF0-051.pdf &Neutrinoless double beta decay in eﬀective ﬁeld theory and simpliﬁed models \\
46 &\cellcolor[HTML]{e2efda}SNOWMASS21-RF5\_RF0-EF8\_EF9-NF2\_NF3-TF7\_TF0-CompF2\_CompF0\_Ruiz\_Richard-118.pdf &\cellcolor[HTML]{e2efda}(NF2) Sterile neutrinos \\
47 &SNOWMASS21-RF6\_RF0-NF2\_NF3-AF2\_AF5-099.pdf &Fixed-Target Searches for New Physics with O(1 GeV) Proton Beams at Fermi \\
\bottomrule
\end{tabular}
} 
\caption{LOIs submitted to the NF02 topical group.}\label{tab:loi}
\end{table}

\cleardoublepage


\renewcommand{\refname}{References}

\printglossary
\bibliographystyle{utphys}

\bibliography{common/tdr-citedb}

\end{document}